\documentclass[twocolumn]{aastex701}

\newcommand\gopreaux{\texttt{GOPREAUX}}
\usepackage{caption}
\usepackage{longtable}

\submitjournal{ApJ}

\begin{document}

\title{GOPREAUX I: Open-source Code and Data to Model Multi-wavelength Emission of Extragalactic Transients using Gaussian Processes}

\correspondingauthor{Craig Pellegrino}

\author[0000-0002-7472-1279]{C. Pellegrino}
\affil{NASA Goddard Space Flight Center, 8800 Greenbelt Road, Greenbelt, MD 20771, USA}
\email{craig.m.pellegrino@nasa.gov}

\author[0000-0001-9227-8349]{T. A. Pritchard}
\affil{University of Maryland}
\affil{NASA Goddard Space Flight Center, 8800 Greenbelt Road, Greenbelt, MD 20771, USA}
\email{tylerapritchard@gmail.com}

\author[0000-0001-7132-0333]{M. Modjaz}
\affil{Department of Astronomy, University of Virginia, Charlottesville, VA 22904, USA}
\email{vru7qe@virginia.edu}

\author[0000-0002-7627-4839]{A. Crawford}
\affil{Department of Astronomy, University of Virginia, Charlottesville, VA 22904, USA}
\email{adrian.crawford@virginia.edu}

\author[0000-0002-1910-7065]{S. Khakpash}
\affil{Department of Physics, Lehigh University, 16 Memorial Drive East, Bethlehem, PA 18015, USA}
\affil{Rutgers University--New Brunswick, Department of Physics \& Astronomy, 136 Frelinghuysen Rd, Piscataway, NJ 08854, USA}
\email{somayeh.khakpash@gmail.com}

\author[0000-0003-1953-8727]{F. Bianco}
\affil{University of Delaware Department of Physics and Astronomy 217 Sharp Lab Newark, DE 19716 USA}
\affil{University of Delaware Joseph R. Biden, Jr. School of Public Policy and Administration, 184 Academy St, Newark, DE 19716 USA}
\affil{University of Delaware Data Science Institute}
\email{fbianco@udel.edu}

\shortauthors{Pellegrino et al.}
\shorttitle{GOPREAUX I: Sample and Modeling Codebase}

\begin{abstract}

Contemporary all-sky surveys have observed thousands of extragalactic transients in the nearby universe, and upcoming surveys will discover exponentially more at higher redshifts. With these large samples, population-level analysis of the photometric behavior of different transient classes is now possible, allowing for photometric classification and physical parameter inference from relatively sparse individual light curves. To enable such studies, we introduce Gaussian process Optimized Photometric Regression of Extragalactic Archival Ultraviolet-infrared eXplosions, a.k.a \gopreaux{}\textemdash a Python package for Gaussian Process Regression of multi-wavelength transient photometry. Our modeling is unique in that it interpolates transient emission across phase and wavelength in a non-parametric, data-driven way. This allows for predictions of light curves and spectra at higher redshifts, where the rest-frame ultraviolet (UV) emission is redshifted into the observer-frame optical or infrared. To this end, we aggregate a sample of almost 1,300 transients observed in the UV and optical with the Neil Gehrels \textit{Swift} Telescope, complemented with additional optical and infrared coverage from surveys such as ZTF and open-source data releases. Our sample includes 275 Type II SNe, 172 stripped-envelope SNe, 72 superluminous SNe, and 58 tidal disruption events, among other classes. Our code and reduced photometry\textemdash comprising over 146,000 photometric observations\textemdash are available as open-source software and data products. Here we discuss our sample criteria, data reduction and modeling methodologies, the multi-wavelength light curves and spectral templates produced by our models, and the future directions in photometric classification and physical parameter inference this code and data repository enables.

\end{abstract}


\section{Introduction} \label{sec:intro}

Time-domain astronomy is currently undergoing a revolution, as current and future optical surveys have begun scanning the sky with unprecedented cadence, depth, and wavelength coverage. These surveys are driving an exponential increase in the number of extragalactic transients discovered each night\textemdash the upcoming Legacy Survey of Space and Time (LSST) at the Vera C. Rubin Observatory alone will detect millions of transients each night \citep{Ivezic2019}. Importantly, these surveys will discover more transients than can be spectroscopically classified, and expand the population of well-sampled transient light curves to higher redshifts. Therefore, photometric classification will be crucial to identifying and following the most interesting objects. This in turn relies on a better understanding of different transients' multi-wavelength, rest-frame behavior to better understand their evolution at higher redshifts (where their emission will be significantly redshifted in the observer frame) and at the cadences expected from Rubin LSST and the \textit{Nancy Grace Roman Space Telescope} \citep{Spergel2015,Akeson2019}.

Thanks to several decades of prolific all-sky surveys\textemdash including ZTF \citep{Bellm2019,Graham2019}, ATLAS \citep{Tonry2018,Smith2020}, and ASAS-SN \citep{Shingles2021}, among others\textemdash as well as multi-wavelength follow-up missions such as the \textit{Neil Gehrels Swift Observatory} \citep{Gehrels2004,Roming2005}, the dataset of transient observations spans orders of magnitude in phase and wavelength. This provides a novel opportunity to begin aggregating complementary observations that cover a more complete range of phases and wavelengths, allowing us to construct transient spectral energy distributions (SEDs) over time. In particular, while two decades of public data from \textit{Swift} have led to a better understanding of the rest-frame ultraviolet (UV) behavior of different spectroscopic classes of transients at a population level \citep[e.g.,][]{Brown2010, Pritchard2014, Vincenzi2019, Hinkle2025}, there is still a need to use these data to better forecast high redshift transient emission. This is to better model the transients that will be observed by upcoming surveys such as Rubin LSST and \textit{Roman}, which have the depth to observe common types of transients at $z \gtrsim$ 0.1, in the optical and infrared.

The recent influx of data has also led to an increased attention toward population-level, data-driven models of many different transient classes. For supernovae (SNe), many works have separately attempted to model the average light curve evolution of different spectroscopic subtypes. Early works \citep[e.g.,][]{Drout2011,Liu2014,Taddia2018} modeled the optical light curves of tens of SNe, including stripped-envelope and core-collapse SNe (for reviews, see e.g., \citealt{Filippenko1997,Modjaz2019}). More recently, these sample sizes have increased to the hundreds, fueled by large collaborations which have released thousands of data points for hundreds of objects, such as the CfA SN Survey for Stripped SNe \citep{Bianco2014}, the Carnegie Supernova Project \citep{Stritzinger2018}, and PanSTARRS \citep{Scolnic2018}, among others. Photometric models are used to generate synthetic photometry of different transient classes, simulate alerts from time-domain surveys, and identify contaminant objects in cosmological samples \citep[e.g.,][]{Hounsell2018}, among other use cases. For example, the Photometric LSST Astronomical Time Series Classification Challenge \citep[PLAsTiCC;][]{Plasticc2018,Kessler2019} simulated millions of transients in LSST bandpasses using a combination of empirical data and analytical models. The Extended LSST Astronomical Time-series Classification Challenge (ELAsTiCC) extended this effort by streaming these alerts via community brokers and quantified their performance \citep{Narayan2023}.

More recently, novel machine learning techniques have been developed to model extragalactic transients. These efforts typically split into two groups\textemdash models to photometrically classify transients, and models to forecast time-series behavior. The former group focuses on classifying transients using survey data, often via supervised or semi-supervised learning methods, including random forest algorithms or neural network architectures \citep[e.g.,][]{Villar2020,Hlozek2023}, while the latter finds parametric or non-parametric representations of ensemble light curve behavior. In many cases the classification efforts rely on some time-series forecasting as well \citep[e.g.,][]{Villar2021}. Among the latter group, Gaussian Process Regression (GPR) specifically has gained traction in time-domain astronomy \citep{Inserra2018,Saunders2018,Vincenzi2019}. GPR is a data-driven model which enables forecasting and uncertainty estimation given an input dataset with arbitrarily many features as well as a choice of covariance, or kernel, function. It assumes each data point is drawn from a Gaussian distribution with an associated covariance and considers joint likelihood between all observed data. When applied to astronomical transients, this allows for Bayesian regression, enabling an interpolation of an object's light curve or SED evolution over time with very few assumptions made about the underlying physics of the system being modeled.

With some exceptions \citep[e.g.,][]{Saunders2018,Vincenzi2019,Boone2019,Qu2021,Hiramatsu2023,Gomez2024}, most GPR applications to astronomical transients in the literature focus on modeling individual filters. While less complex and more computationally tractable, this method loses critical information about filter dependency by effectively interpolating each light curve independently from one another. In principle, however, there is inter-filter dependency for most transients, depending mainly on the physical mechanisms powering the emission of different transient classes. Additionally, using cross-filter behavior to forecast emission is imperative when extrapolating light curve and SED behavior at higher redshifts, when the rest-frame UV and optical is redshifted into the IR. Therefore, accounting for rest-frame emission across as wide a wavelength range as possible when modeling their full SED evolution is crucial to photometrically identifying transients in current and future time-domain surveys.

While the sample sizes have grown exponentially and more advanced classification and regression models have been developed, the field of extragalactic time-domain astronomy lacks a cohesive framework that ties together a contemporary data archive with data-driven, nonparametric spectrophotometric modeling. Specifically, to date no individual work has done all of the following:
\begin{itemize}
    \item model multi-dimensional spectrophotometric data using GPR;
    \item produce a uniform set of time-evolving SEDs, interpolated over a large range of wavelengths and phases, for all major spectroscopic transient classes;
    \item provide a catalog of astronomical data from the ultraviolet to infrared to serve as a gold standard for modeling efforts; and
    \item develop open-source software to enable data-driven modeling for an arbitrary collection of transients, complete with routines to transform raw data into actionable models in a classification-agnostic manner.
\end{itemize}

Here we introduce a novel software package that implements GPR for multi-dimensional transient light curve data. Our software package models full transient SED by forecasting across wavelength and phase simultaneously, producing model SED ``surfaces" that give an interpolated SED at arbitrary phase, or an interpolated light curve at arbitrary wavelength. We use this fitting to produce template model SED surfaces for common transient classes, focusing mainly on SNe but also modeling TDE behavior as well. The codebase contains one of the largest samples of extragalactic transient data assembled, with more than 146,000 photometric detections of almost 1,300 objects, from the UV to IR, with \textit{Swift} as the backbone of the sample. These software and data are made publicly available and open-source in an easily installable and user-friendly Python package, named \texttt{GOPREAUX}\textemdash Gaussian process Optimized Photometric Regression of Extragalactic Archival Ultraviolet-infrared eXplosions. 

This work describes the \gopreaux{} software, data sample, and data models, while we reserve an in-depth comparison of the results to an upcoming work. This paper is organized as follows: in Section \ref{sec:gpr}, we overview GPR and its application to time domain astronomy today. In Section \ref{sec:methods}, we describe our sample of transient data and our data reduction techniques. We outline our codebase and modeling routine in Section \ref{sec:code} and detail the outputted model objects in Section \ref{sec:fits}. Finally, we discuss future directions and conclude in Section \ref{sec:discussion}.

\section{Gaussian Process Regression Description}\label{sec:gpr}

In this Section, we give a brief overview of Gaussian Process Regression (GPR) and its utility in astronomical contexts. For a comprehensive overview of Gaussian Processes, see \cite{Rasmussen2006}, \citet{ForemanMackey2017}, and \citet{Aigrain2023}. In short, Gaussian Processes are an extension of the Gaussian distribution to functions. In the Gaussian Process formalism, functions can naively be considered infinite-dimensional vectors. By considering only their values at a finite set of points, Gaussian Processes allow for regression across the entire domain of these functions. This allows for modeling without any knowledge of the functional form of the input data and naturally offers uncertainty estimates by extending a Gaussian distribution to each coordinate of the function. 

GPR have seen an uptick in use in astronomical research and data modeling over the past several years, particularly in time-series data forecasting. For a comprehensive overview, see the discussion in \citet{Aigrain2023}. In short, many of these use cases have centered on stellar astrophysics and exoplanet research\textemdash for example, analyzing stellar periodicity \citep[e.g.,][]{Brewer2009,Angus2018,Boettner2025} or exoplanet transit signals in high-cadence stellar light curves \citep[e.g.,][]{ForemanMackey2014}. More recently, several works \citep[e.g.,][]{Boone2019,Saunders2018,Qu2021,Hiramatsu2023,Khakpash2024,Gomez2024} have used GPR to model extragalactic transient light curves and spectra. 

In time-domain astrophysics, GPR have several key advantages. The first is that they model light curves or multi-wavelength emission in a data-driven manner without assuming a functional form of the data being fit. Traditionally, transient observations have almost always been modeled using an analytical model rooted in the underlying physics powering the transient emission. For example, \citet{Arnett1982} is a common formalism used to model the light curves of SNe powered by radioactive decay of $^{56}$Ni. Other similar models exist for an array of other powering sources, such as circumstellar interaction \citep[e.g.,][]{Chevalier1982} and magnetar spin-down \citep[e.g.,][]{Kasen2010,Woosley2010}, among others. Unlike in these cases, GPR is agnostic to the physical mechanisms powering the transient emission, providing a model that is not biased by analytical formalism or researchers' assumptions.

The second advantage of using GPR in time-domain astrophysics is their innate flexibility\textemdash they naturally account for sparse or sporadically-sampled data and can be generalized to arbitrary dimensions. This complements the nature of transient astronomy nicely. For example, data may be taken at a higher cadence in one filter and a sparser cadence in another. However, with multi-dimensional GPR, we can predict the behavior of the sparsely-sampled light curve using the observed behavior of the other. This allows us to combine sparse data sets from different instruments, filters, and even different objects to yield a GPR likelihood for any type of transient.

A key facet of GPR that enables this is the choice of covariance, or kernel, function. Each kernel function is parametrized such that key data features are emphasized. For example, the commonly-used Radial Basis Function (RBF) kernel is parametrized by the distance between input points and is infinitely differentiable, enforcing smoothness in the output GPR object. 

Kernel ``hyperparameters" are optimized during the regression by maximizing a marginal likelihood function, which differences the residuals between the observed data and a mean function and the kernel function values, to construct a posterior distribution. The mean function in this case will be taken to be zero (see Section \ref{sec:code} for more details). Finally, the predictive distribution of functional behavior across the parameter space is a Gaussian distribution, calculated by averaging over all parameters, weighted by the posterior distribution.

In short, GPR excels at modeling complicated, multi-dimensional data without input as to its functional form or behavior. Because of this, it naturally allows for uncertainty estimation and extrapolation. On the other hand, this comes at a computational cost, as most implementations of GPR run in $\mathcal{O}(n^3)$ time, though some implementations have been adapted to be more computationally performant \citep{Ambikasaran2015,ForemanMackey2017,ForemanMackey2018}. However, for our sample size and purposes, this trade-off is worth the predictive behavior and sophisticated multi-dimensional modeling of transient SED evolution.

\section{Sample Description and Data Reduction}\label{sec:methods}

\begin{figure*}
    \centering
    \includegraphics[width=0.95\linewidth]{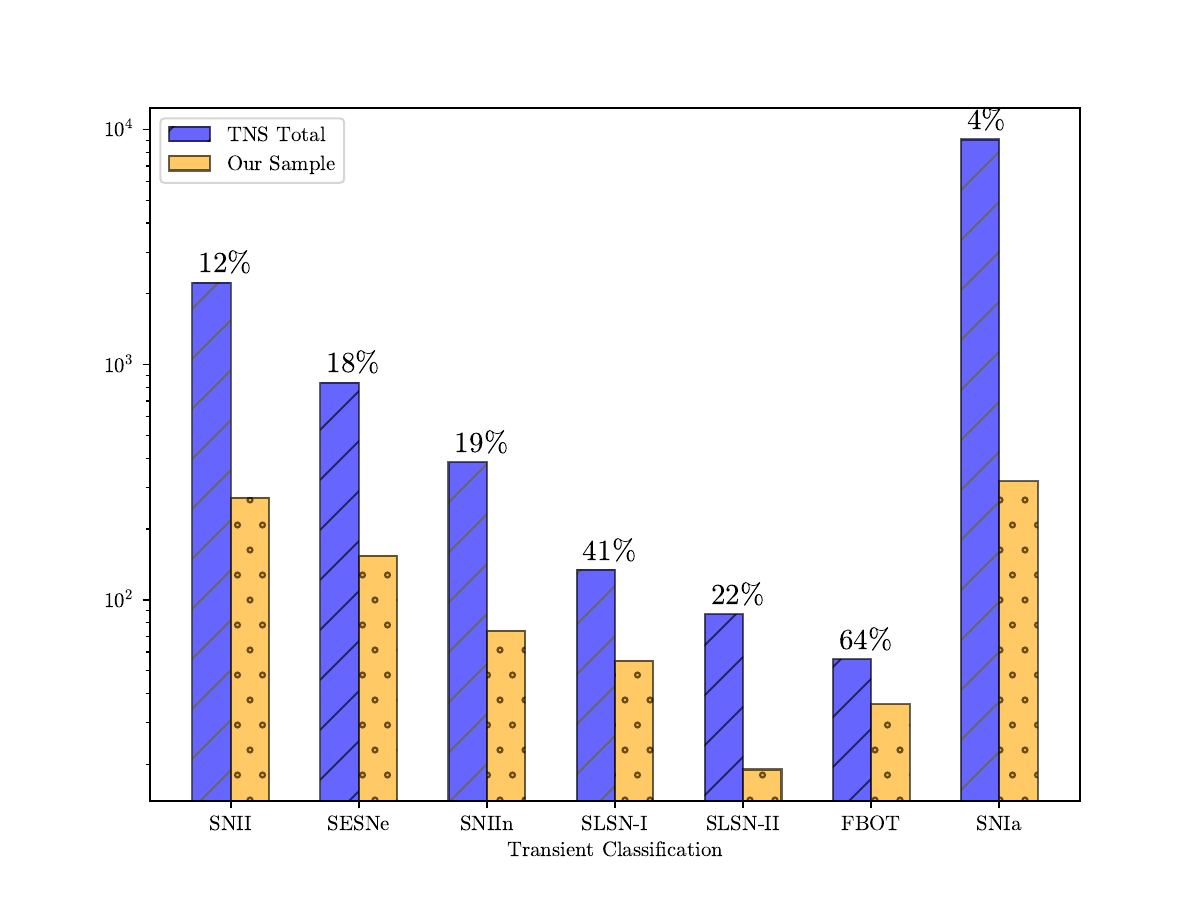}
    \caption{Bar chart comparing the number of transients in each spectroscopic class in our sample (dotted orange bars) to the total number number of transients spectroscopically classified on the Transient Name Server (striped blue bars). Note that the y axis is in logarithmic scale. The percentage of the TNS objects in our sample is given above each bar.}
    \label{fig:objectspertype}
\end{figure*}

The first aim of this work is to produce a curated and comprehensive UV-to-NIR dataset of transients, including all supernova subtypes and other common transients with similar phenomenology. These data will not only enable the development of our own \gopreaux{} software, but by making it public, we provide a resource to the community that will support further studies. Including UV photometry enables explorations and predictions relevant to the upcoming LSST survey, which will significantly expand the detection rate of transients at high redshift (e.g. $>1,000$ SNe Ia at $z > 0.8$ \citealt{2024ApJS..275...21G}).

Our \gopreaux{} software repository contains both the codebase to construct GPR models as well as a broad library of transient data for the fitting. In this Section, we describe the data and the sample of objects the data are drawn from, as well as our uniform data reduction processes and sample filtering criteria used to ensure our final fitting routines are robust and valid. This work has only recently become possible thanks to the exponential increase in publicly available data, as missions such as \textit{Swift} enter decades of workhorse use, and ground-based time domain surveys release millions of publicly available data points over their lifetimes. 

We model and generate templates for the most common extragalactic transient spectroscopic classifications. Our efforts focus on modeling different spectroscopic classes of core-collapse SNe; however, we include in our sample any astrophysical transients reported to the Transient Name Server (TNS) with \textit{Swift} observations. Our sample is broken down into seven main classifications:
\begin{enumerate}
    \item SN II
    \item Stripped-envelope SN (SESN)
    \item SN IIn
    \item Type I Superluminous SN (SLSN-I)
    \item Type II Superluminous SN (SLSN-II)
    \item FBOT
    \item Other
\end{enumerate}
These classes are further broken down as follows: within the Stripped-envelope SNe class, we consider Type Ib, Ic, and IIb SNe; within the FBOT class, we consider Type Ibn and Icn SNe and luminous fast blue optical transients; and within the Other class, we consider Type Ia SNe and related subtypes, Tidal Disruption Events, and unclassified transients.

\begin{deluxetable}{ccccc}[t!]
\tablecaption{Sample Breakdown \label{tab:sample}}
\tablehead{
\colhead{Type} & \colhead{$\#$ Objects} & \colhead{$\#$ Objects w/} & \colhead{$\#$ Detections}  & \colhead{$\#$ Non-} \\
\colhead{} & \colhead{} & \colhead{Measured Peak} & \colhead{} & \colhead{detections}}
\startdata
SN II & 275 & 204 & 44,585 & 13,781 \\
SESN & 172 & 103 & 16,817 & 5049 \\
SN IIn & 72 & 56 & 10,044 & 2997 \\
SLSN-I & 54 & 50 & 10,380 & 3443 \\
SLSN-II & 18 & 17 & 3322 & 760 \\
FBOT & 31 & 31 & 2625 & 3297 \\
Other & 695 & 347 & 58,301 & 17,016 \\
\enddata
\end{deluxetable}

\subsection{Sample Criteria}

We limit our sample to all publicly-reported transients observed by \textit{Swift} through August 2023 to ensure that all transients we fit have at least UV-optical coverage. \textit{Swift} photometry is supplemented with other publicly-available data sources, including data from time-domain surveys, specifically ZTF, ATLAS, and ASAS-SN; data on the Open SN Catalog \citep{Guillochon2017}; and public data releases from research collaborations. Classifications, coordinates, and redshifts were retrieved from the TNS and Wiserep \citep{Yaron2012}, two publicly available databases of transient discovery and classification information. The ingested classifications were vetted by comparing to any existing classifications in the literature, and any obvious incorrect classifications were adjusted. For more discussion on our classification verification, see Section \ref{subsec:typing}. 

\begin{figure*}    
    \centering
    \includegraphics[width=0.90\linewidth]{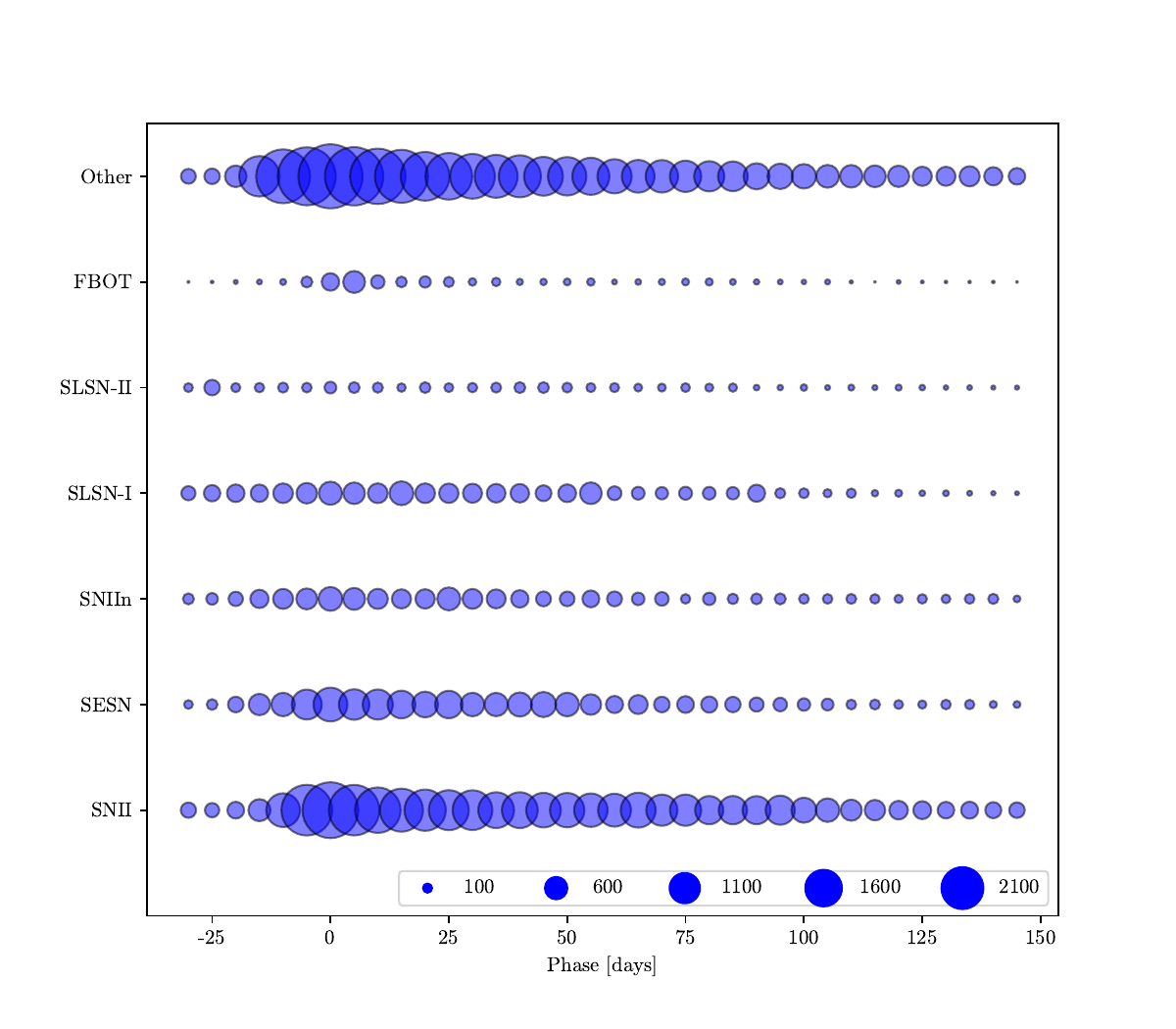}
    \caption{The number of individual detections as a function of phase for the transients in each major spectroscopic class in our sample. The size of each point scales with the number of detections in that phase bin. Each bin covers 5 days. There are a total of 146,074 detections across all transient classes in our sample.}
    \label{fig:detsperphase}
\end{figure*}

In total, our sample consists of 1,297 individual objects across 19 distinct spectroscopic classes. This represents one of the largest ensembles of transients ever processed and modeled in a uniform way, particularly in the UV, which historically has lacked uniform and comprehensive modeling efforts. Multiwavelength modeling, including recent works, generally focuses on specific subtypes. For example, \citet{Hiramatsu2024} recently assembled a large sample of SNe IIn and SLSN-II, modeling their multi-wavelength light curves using multi-dimensional GPR. \citet{Gomez2024} presented a similar sample of SLSN-I, again modeled using a GPR implementation. \citet{Khakpash2024} (hereafter K24) used GPR to create UV-optical light curve templates of stripped-envelope SNe (SNe Ib, Ic, and IIb) using 165 objects. They establish procedures and methodologies for the use of GPR to fit light curves of transients, many of which we inherit and leverage (see Section \ref{sec:gpr}). While these works may include more individual objects or detections in their respective samples, our overall sample size is larger, we apply our modeling uniformly across all spectroscopic classes. A more detailed breakdown of the number of objects, detections, and non-detections per spectroscopic class is given in Table \ref{tab:sample}. A complete description of each object in our sample is given in Table \ref{tbl:sampleproperties}.

Our sample size compared to the total number of objects on the TNS is shown in Figure \ref{fig:objectspertype}. Our sample contains a significant fraction of all spectroscopically classified transients from each major class. Overall, our dataset contains a total of 146,074 detections and 46,343 nondetections that we use in our fits\textemdash this only counts photometry from transients with well-defined light curve peaks, redshifts, and coordinates (for more information on object validation and data pre-processing, see Section \ref{sec:code}).

\subsection{Data Collection and Reduction}

\begin{figure*}    
    \centering
    \includegraphics[width=0.95\linewidth]{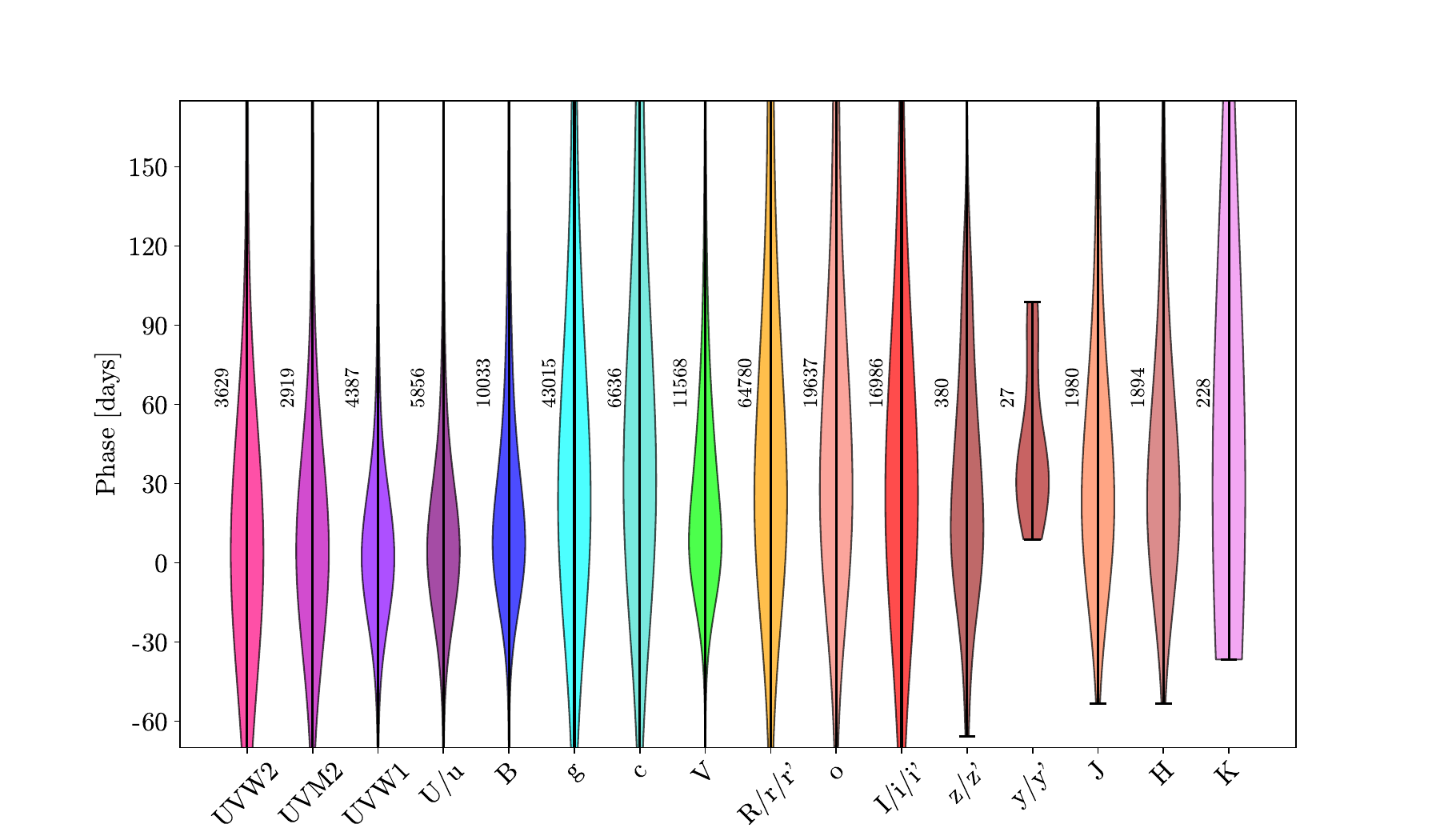}
    \caption{Violin plot showing the distribution of detections per filter as a function of phase for the entirety of our sample. The number of detections in each filter is shown to the left of each distribution. While each filter has a greater number of detections around peak brightness, we still maintain excellent coverage from the UV to the IR at very early and very late phases.}
    \label{fig:detsperfilter}
\end{figure*}

All data for the objects in our sample are publicly available via GitHub.\footnote{\url{https://github.com/crpellegrino/gopreaux}.} Figure \ref{fig:detsperphase} shows the number of detections as a function of phase for each transient class, while Figure \ref{fig:detsperfilter} shows the distribution of detections as a function of phase across filters. Here we describe the individual efforts to process and aggregate data from these different sources:

\subsubsection{Swift Sample and Reductions}\label{subsec:swift}

Swift aperture photometry was calculated using a custom UVOT pipeline, based off work by \cite{Brown2014} with current calibration files and zeropoints \citep{Breeveld2011}. Transient flux was calculated using a 3$\arcsec$ aperture centered on the transient location. Additionally, our implementation differs from that of \cite{Brown2014} in that we automatically performed background and host galaxy subtraction. Background subtraction was performed by measuring the flux in a larger aperture centered on regions with no sources, identified using \texttt{sep} \citep{Barbary2016}, a Python implementation of Source Extractor \citep{Bertin1996}. For host galaxy subtraction, template images either before explosion or after the transient flux had faded were selected programmatically. For templates after explosion, we filter for images taken more than 6 months after the time of first detection, or more than a year for longer-lasting transients such as TDEs and SNe IIn. All photometry was calculated in Vega magnitudes and later converted to AB magnitudes using the conversions given by \citet{Siegel2010}.

To validate our reductions, we compare our calculated UVOT photometry with publicly-available data. We primarily utilize the Open SN Catalog and the open-source SOUSA database of UVOT reductions \citep{Brown2014}. We find that across all of our classes, 95\% of objects have fewer than 6\% of data points that differ at the 5$\sigma$ level. In cases of discrepancy, we choose to use our calculated values. This is done to ensure that all the objects in our sample contain uniformly-processed data, as our archive of reductions contains many more objects than are currently publicly available. 

\subsubsection{Time-domain Survey Data}\label{subsec:surveys}
\begin{figure*}
    \centering
    \includegraphics[width=0.9\linewidth]{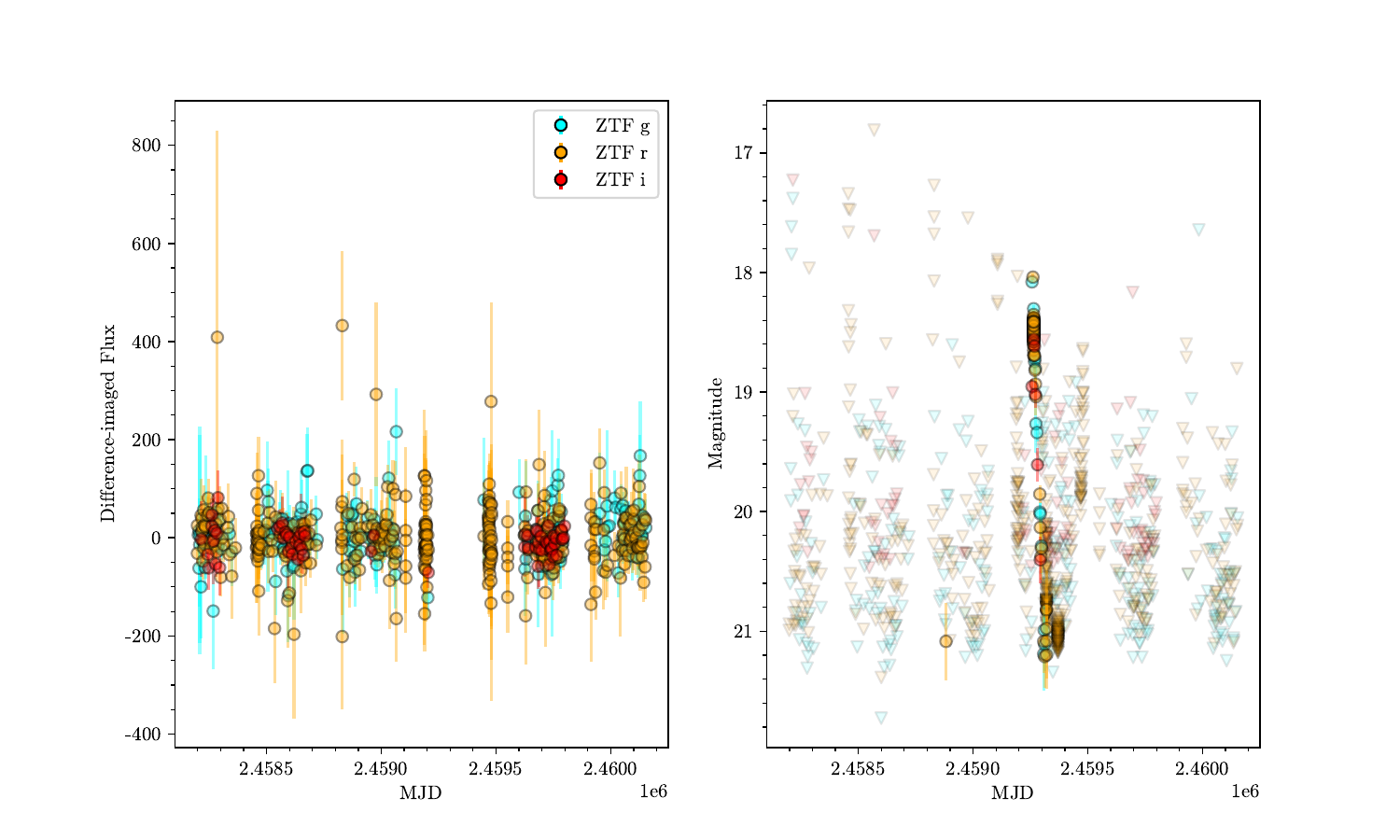}
    \caption{Left: Forced photometry from difference images retrieved from the ZTF forced photometry server. Flux values in each filter-camera combination are used to establish a flux baseline. Right: The final magnitude-calibrated ZTF photometry for the same object. 5$\sigma$ upper limits are shown in lightly-shaded triangles while detections are shown in colored circles.}
    \label{fig:ztf}
\end{figure*}

Swift UV and optical data are complemented with publicly-available data for those SNe. We use optical data from ground-based surveys including ZTF, ATLAS, and ASAS-SN. Forced photometry was requested from the ZTF \citep{Bellm2019,Graham2019}, ATLAS \citep{Tonry2018,Smith2020}, and ASAS-SN \citep{Shingles2021,Hart2023} forced photometry servers, which provide difference-imaged photometry. In all cases, anomalous detections and subtraction artefacts were identified and removed using the following procedure: first, a baseline flux level was established using the raw count measurements from each telescope-filter combination. This baseline measurement was subtracted from the measured counts from each exposure. Next, we identify detections using the signal-to-noise ratios (SNRs) for each measurement. We use detections with SNRs $\geq$ 3 and identify others as non-detections at the 5-$\sigma$ level. Finally, these baseline-subtracted flux measurements were converted to AB magnitudes, along with their associated uncertainties. An example of raw and reduced data are shown in Figure \ref{fig:ztf}.

This process yields final, publication-ready data. All steps replicate those outlined by various forced photometry servers, such as the ZTF FPS.\footnote{\url{https://ztfweb.ipac.caltech.edu/cgi-bin/requestForcedPhotometry.cgi}.} To verify that our results are consistent with those published in the literature, we compare published objects on the Open SN Catalog with reductions in our codebase. We find that 95\% of objects have only 10\% or fewer detections that are inconsistent with published values within the uncertainty ranges. As with our \textit{Swift} UVOT reductions, when there is a discrepancy, we choose to use our reduced values for consistency among the objects in our sample. Given the various other pre-processing and processing steps outlined in Section \ref{sec:code}, we are confident any systematic differences in reduction routines will not significantly impact our final SED template models.

\subsubsection{Open-source Data Releases}\label{subsec:opensourcedata}
We also include published data that is publicly available via the Open SN Catalog. These data on average cover a large wavelength and phase range, particularly at filters such as the NIR, which are less frequently sampled by all-sky surveys. Open SN Catalog data were retrieved using their public-facing API \citep{Guillochon2017}. Data were queried for all transients in our sample, and the retrieved data was cross-matched with our existing sources (e.g., Swift, ZTF, etc.) to check for duplicates. Any duplicate Open SN Catalog data was discarded, while that data not already in our sample was saved and used in our fitting routines.

Finally, we supplement our NIR datasets with other published samples from transient research collaborations. We include the optical and NIR light curves of SESNe from \cite{Bianco2014} to add to the NIR coverage of our SESNe population\textemdash a main focus of our modeling and future research efforts (see Section \ref{sec:discussion} for more detail). These data cover the \textit{U}-band to the NIR in \textit{JHK}-bands via the Whipple Observatories' optical and NIR telescopes operated by the Harvard Smithsonian Center for Astrophysics. Template subtraction was performed for all objects with host contamination. For more information, see \cite{Bianco2014}.

\subsection{Photometric Corrections}

Once the raw photometry has been reduced for the transients in our sample, few corrections or post-processing steps are done. This is to ensure that our resulting models are as data-driven as possible, without excessive assumptions about the underlying nature of the transients themselves. For example, we do not make an effort to correct for extinction from the host galaxies of the transients. While host galaxy line-of-sight extinction may redden transients' SEDs, potentially reducing the observed brightness by magnitudes in the UV and bluer filters, correcting for this effect is both cumbersome and difficult to do consistently across the hundreds of transients in our sample. Instead, we consider host extinction as part of the natural spread in the range of colors and magnitudes of a given transient class, driven in part by secondary factors such as explosion site preference and host galaxy characteristics which contain some features of the observed data.

On the other hand, we do correct all photometry for Milky Way extinction. This is because Milky Way dust extinction is well understood and can be applied to any future transient easily, without any knowledge of the properties of the transient or its host galaxy. To correct for Milky Way extinction in a uniform way, we use a Python implementation \citep{Green2018} of the dust maps of \citet{Schlegel1998}. Filter-dependent extinction corrections are determined using the effective wavelengths of each filter, which are calculated ad hoc for each photometric data point (see Section \ref{sec:code} for more details). The extinction-corrected photometry for each transient is then saved as the final, processed data archive to be re-initialized during the fitting and modeling process.

Finally, we do not calculate $k-$corrections for our photometry. As we limit our sample to objects observed by \textit{Swift}, we are preferentially modeling low-redshift transients, for which $k-$corrections are trivially small. Additionally, proper $k-$corrections across the diverse range of spectroscopic classes in our sample would require prior knowledge of their spectral evolution and inherent color\textemdash exactly what we are seeking to model. Therefore, we choose to ignore $k-$corrections to avoid biasing our modeling efforts.

\subsection{Spectroscopic Classification Determination}\label{subsec:typing}

Our sample of transients is organized by spectroscopic class and subclass in order to inform our modeling efforts. However, in some cases, transient classification can be ambiguous or variable. The values reported on the TNS, which are often determined using a single spectrum within days of explosion, may be incorrect or change with time. This is particularly true of stripped-envelope SNe, for a number of reasons: the diagnostic features in the spectra of these objects may be obscured by other features, slow to develop, or are strongly wavelength dependent \citep[e.g.,][]{Williamson2023}. Additionally, fewer template spectra of these objects exist \citep[e.g.,][]{Yesmin2025}, which may cause classification codes such as SNID or Superfit \citep{Howell2005,Blondin2007} to incorrectly associate them with other objects. Finally, their light curve shapes around peak brightness are often similar to each other and to other classes of transients, removing an additional diagnostic which often aids in spectroscopic classification effort \citep[e.g.,][]{Khakpash2024}.

To ensure the spectroscopic classifications from the TNS are as accurate as possible, we programmatically search for re-classifications in published works for the objects in our sample. We utilize the Astrophysics Data System (ADS) API and query for refereed works that contain the name of each transient in our sample in the title, abstract, or keywords. We visually inspect the returned articles, focusing on those that suggest a classification that does not agree with the TNS value. In cases of a clear re-classification (such as strong matches to template spectra), we change the classification in our repository. These values are reflected in Table \ref{tab:reclassifications} along with the associated references. On the other hand, when there are discrepancies or disagreeing values, we mark these transients as ``Transitional'' rather than deciding on a classification. While a systematic, uniform classification effort across our entire sample, using spectral classification codes, may lead to more robust classifications, such an effort is outside the scope of this work. Additionally, it may also result in additional ambiguous classifications owing to the inherent uncertainty in transient spectroscopic diversity and the limitations of spectral template libraries. By making our reclassification efforts relatively narrow in scope, we are in turn capturing a broader range of behavior for the objects in each type, which is important for covering the full spread in a class's photometric behavior and evolution. Therefore, we follow a relatively conservative approach toward our sample classifications by adhering to published values.

\subsection{Final Sample Determination}

Our final sample is defined by the above processing and filtering criteria. We require a transient to have the following:

\begin{itemize}
    \item known and fixed coordinates, to enable Milky Way extinction correction,
    \item known and well-determined spectroscopic classification and redshift,
    \item well-sampled photometry, particularly around the light curve peak, to enable peak phase and magnitude determination.
\end{itemize}

Any transient that does not meet these criteria is excluded from the fitting and modeling routines. In particular, ``Transitional'' objects (see Section \ref{subsec:typing}) are not included in the samples used to make our final template models. This is to avoid contamination from misclassified objects, ensuring that our templates are as reflective of the underlying nature of each class as possible. 

\section{\texttt{GOPREAUX} Data Processing and Modeling Description}\label{sec:code}

\begin{figure*}
    \centering
    \includegraphics[width=0.9\linewidth]{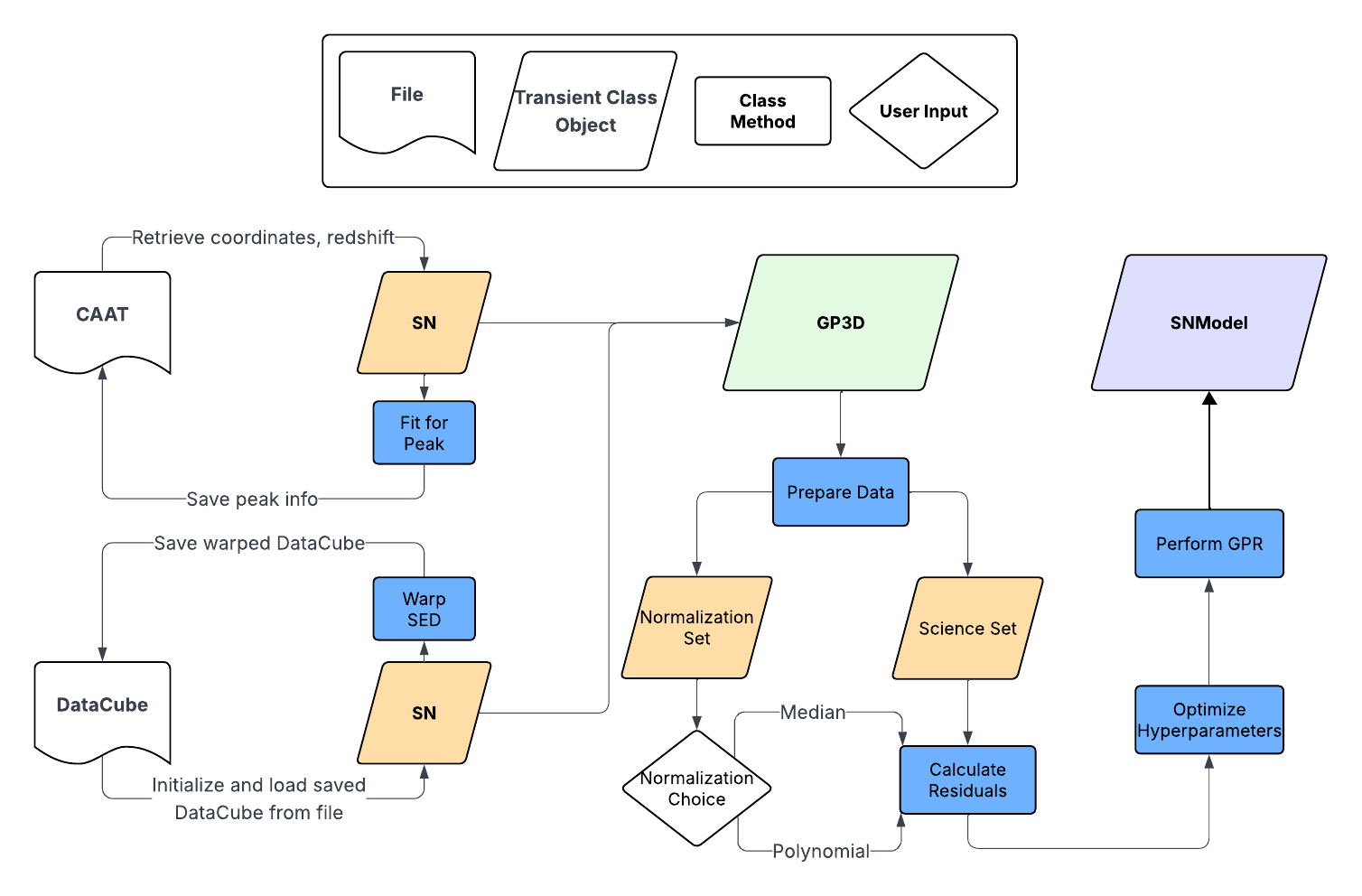}
    \caption{A flow chart demonstrating the \gopreaux{} workflow, including pre-processing, data preparation, and modeling routines. The legend at the top details the meaning of each shape in the chart. In particular, Python class objects are shown as parallelograms with names in bold, and functions (methods) on those classes are shown as rectangles.}
    \label{fig:flowchart}
\end{figure*}

\gopreaux{} is written in Python and available for download and use via GitHub.\footnote{\url{https://github.com/crpellegrino/gopreaux}.} It is self-contained and requires little specialized knowledge to access and use the template model objects; however, users with programming experience and interest in modifying the code are empowered to create their own templates and model architecture by modifying the codebase or defining new samples to model. The open-source codebase contains both the data used in the fitting as well as the code routines to aggregate the data, perform necessary pre-processing steps, and perform the GPR. These steps are demonstrated graphically in Figure \ref{fig:flowchart}. Users are also able to add new objects and data to fit as well. Here we outline the main steps involved in pre-processing, processing, and modeling of the data sample.

Information on the transients in the codebase, including classification, coordinates, redshift, and time and magnitude of peak brightness, are aggregated in a centralized object called a Catalog of Archival Astronomical Transients (CAAT). At a base level, the CAAT is a \texttt{.csv} file containing a row for each transient. This data can be instantiated and accessed using the \texttt{CAAT} Python object, which enables adding transient information in a programmatic way; for example, redshifts and coordinates can be cross-matched and added from a TNS \texttt{.csv} file, and peak brightness information can be added in an interactive, programmatic way, as described below.

After downloading and saving the data for our transients, the primary pre-processing step is to estimate the date of and magnitude at peak brightness. This is to standardize the transient light curves in a way that allows for uniform fitting of objects across different spectral types and distances. Unlike other modeling efforts, we are fitting across wavelengths as well as phases simultaneously, so we must pick a single wavelength or filter to use as the reference for peak brightness. We choose to use the filter closest to the wavelength range 4,000 -- 5,000 \AA\ in the rest frame, as this wavelength region is covered by \textit{Swift} as well as most ground-based all-sky surveys, ensuring ample coverage for most of the objects in our system. We exclude from the remainder of the modeling process transients for which we could not constrain the time and magnitude of peak in any of these blue filters. All photometry in each filter is then shifted relative to the measured time and magnitude at peak. 

To fit for peak time and brightness, we follow a similar procedure to that outlined in \citet{Bianco2014}. First, extinction-corrected photometry within thirty rest-frame days of peak brightness, determined either from the local minima of the light curve or from a user-supplied initial value, is fit with a second degree polynomial. To estimate uncertainty in phase and magnitude, the number of photometry points being fitted around the peak is iteratively varied by randomly changing the range of phases being fit (by [-3, 3] days for our calculated values), and the time and magnitude of peak from each iterative fit is recorded. The final estimate and its uncertainty are given by the mean and standard deviation of the reported values, respectively. 

\gopreaux{} enables this process to run either automatically or interactively with user inputs for the filter being fit, an estimate of the time of peak brightness, and feedback to redo fits with broader or narrower fit parameters depending on the quality of the reported fit. For the objects in our sample, we ran the fitting interactively to qualitatively assess each fit. We manually chose the filter closest to the 4,000 -- 5,000 \AA{} range with the densest sampling around the light curve peak. In some cases, the automatically chosen light curve peak had to be manually adjusted\textemdash for example, if there was a spurious detection brighter than the true peak. We paid careful attention to SNe whose light curves show double-peaked behavior, such as some Type IIb SNe. For these objects, we manually identified the secondary peak powered by radioactive decay of $^{56}$Ni, and not the earlier shock-cooling powered peak, to standardize their behavior with other classes of SNe.

\begin{figure}
    % \centering
    \hspace{-1cm}
    \includegraphics[scale=0.65]{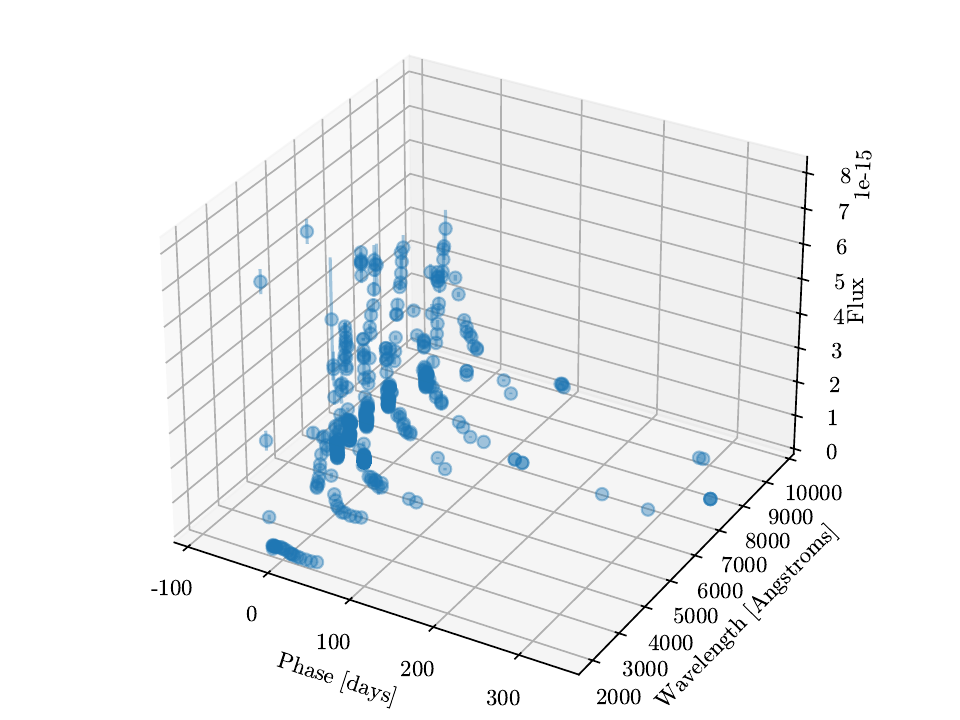}
    \caption{The final ``data cube" used as input to the GPR fitting routines. Each point represents a photometry measurement in a given filter. During pre-processing, the effective wavelengths of each measurement are shifted to the transient rest frame and then ``warped" iteratively until the measured photometry matches the filter functions convolved with the interpolated SED at each epoch.}
    \label{fig:datacube}
\end{figure}

The second pre-processing step necessary for the GPR modeling is to convert a transient's photometry from magnitudes to fluxes at a certain wavelength. This is done to standardize photometry across filters, which may have different zeropoints and magnitude systems, and to effectively utilize nondetection information in a physically meaningful way. The simplest approach is to effectively ``pin" the flux in each filter at that filter's central wavelength in the transient's rest frame. However, assuming all the flux in a filter is at a single wavelength can produce unphysical interpolated SED shapes which, when convolved with the filter function, yields inaccurate synthetic photometry. This is particularly important for the Swift UV filters, which have long red tails in their transmission curves which can bias our flux measurements to bluer wavelengths. 

To mitigate this issue, we introduce an iterative procedure for ``warping" the interpolated SED until it converges to match the observed photometry. First, we linearly interpolate between fluxes at all phases that have observations in at least 3 filters within a day of each other. This process creates an interpolated SED. Next, the interpolated SED is iteratively warped by shifting the effective wavelengths of each filter until the SED convolved with the filter function (a ``synthetic" flux measurement) matches the measured flux in that filter. The filter curves and effective wavelength measurements were taken from the SVO filter profile service \citep{Rodrigo2012,Rodrigo2020}. For each iteration, the wavelengths are shifted proportional to the difference between the measured and synthetic flux values in each filter (e.g., larger differences lead to larger shifts). Once these flux values agree to within 5\%, we then save the shifted effective wavelengths for those photometry points and use them as the effective wavelength values in our GPR fitting. As a result of this process, and because of the transients' distribution across redshifts, we effectively sample a larger range of discrete wavelength values. This helps in the GPR interpolation by augmenting our training set in wavelength space. An example of this final warped ``data cube" is shown in Figure \ref{fig:datacube}.

Nondetections are included in this warping step and are given errors of 0.1 mag to inform the GPR. However, this comes at the cost of the model fits sometimes behaving unphysically to conform to nondetection values. To remedy this, during our pre-processing steps we eliminate any nondetection information which is not more constraining than real detections. This includes nondetections that are at a higher flux value than nearest detection information. We also discard any nondetections between the first and last detection as well as any nondetections outside this range that are less constraining than the nearest nondetection to the first and last detection.

Fitting is performed in logarithmic space for both phase and wavelength values. This has several advantages (as described in K24), the primary of which is to mitigate some of the nonstationary behavior inherent in SN SED evolution. Our implementation of Gaussian Process regression assumes that kernel parameters are stationary, or unchanging with time. However, SN light curve behavior is inherently time- and wavelength-dependent\textemdash flux at earlier phases and in bluer wavelengths evolves more rapidly than flux in redder filters or at later phases, primarily due to the rapid cooling of the SN photosphere as well as intrinsic differences in SN powering mechanisms at early phases, which tend to be shorter lived. Transforming phases and wavelengths into logarithmic space effectively removes much of this time dependence. In our fitting routine, user inputs are allowed for defining the minimum phase, used as the baseline in the logarithmic conversion. For example, users may choose to model phases beginning at -20 days relative to peak brightness, and set the \texttt{log$\_$transform} flag equal to 22, effectively ``stretching" the light curve by two days at early phases.

\begin{figure*}
    \centering
    \includegraphics[width=\linewidth]{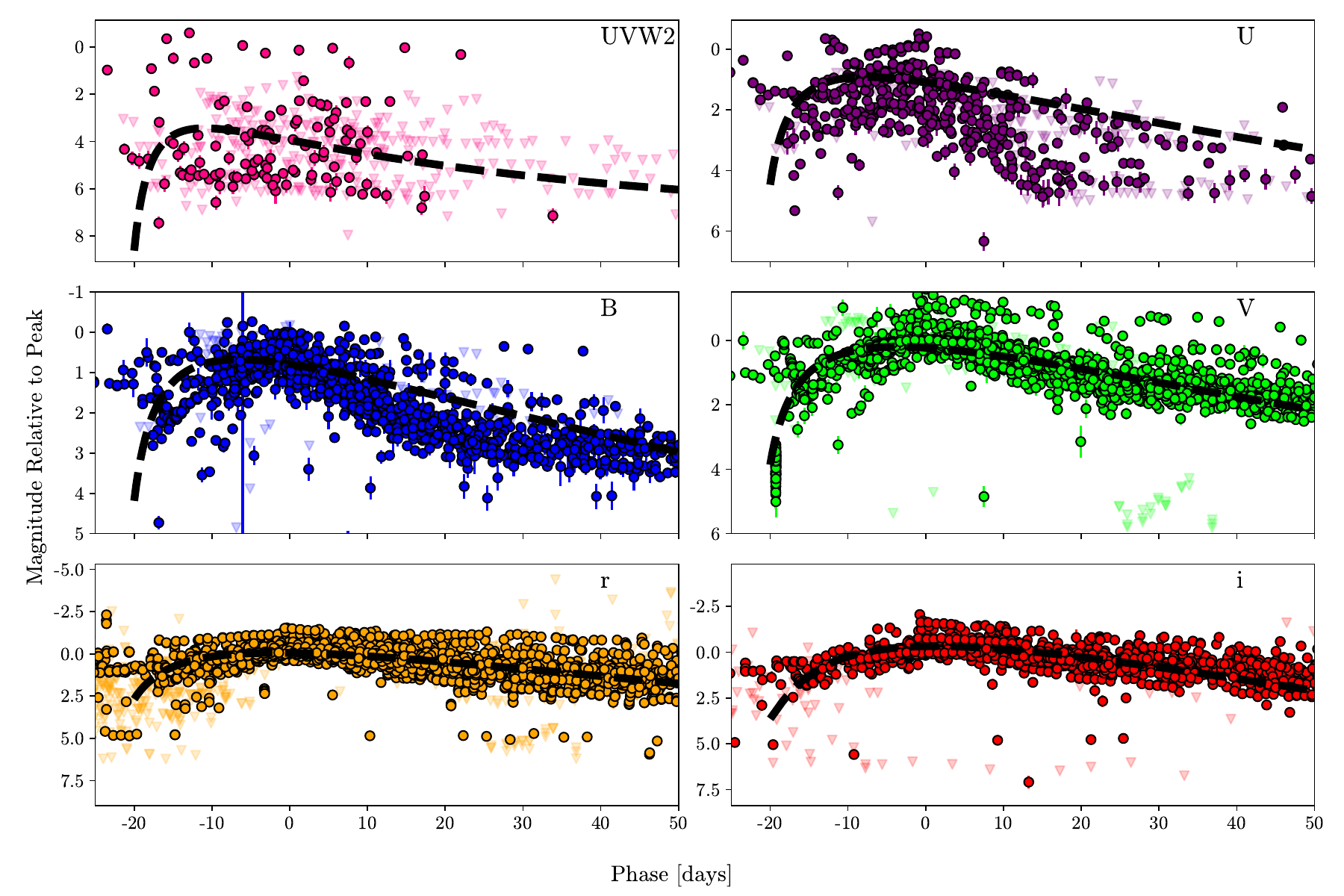}
    \caption{Polynomial templates (dashed black lines) calculated from all stripped-envelope SNe detections (solid points) and nondetections (shaded triangles) in six representative filters. Filter names are given in the top right of each plot. The polynomial template is subtracted from the photometry of each SN in the fitting sample to calculate photometric residuals, which are then fit as part of our GPR routine. This is to approximate a mean zero data set; therefore the polynomial fits are not meant to exactly capture the ``typical'' stripped-envelope SN behavior, but rather the peculiar and specific behavior of each SESN. This procedure is applied separately to all classes in our sample (see Table \ref{tab:sample}).}
    \label{fig:polynomial}
\end{figure*}

Performing the Gaussian Process regression and model creation is done through the custom \gopreaux{} \texttt{GP3D} class, which utilizes the \texttt{scikit-learn} \citep{Pedregosa2011} \texttt{GaussianProcessRegressor} functionality. \texttt{GP3D} is initialized with a collection of transients to model as well as a collection of transients to compare against. As discussed in K24, comparing the photometry of the input transients to another collection of objects is essential to our GPR implementation. This is because GPR assumes a mean-zero function by default. To make this assumption valid, we must subtract off a ``template'' light curve for each transient, to then let the GPR fit the photometric residuals at each epoch. Similarly to what was done in K24, we construct a simple template by aggregating all objects in the class sample (e.g. SESN) and modeling the aggregate light curves with either a polynomial or a median function. This creates a template SED surface which is subtracted from the SEDs of the objects to fit (the ``Science Set" in Figure \ref{fig:flowchart}) at each time point. An example of a polynomial template is shown in Figure \ref{fig:polynomial}. This process is applied separately to all classes in our sample (see Table \ref{tab:sample}) automatically, with minimal user input or supervision required, and is demonstrated in Figure \ref{fig:fits} (top right).

The transient residuals, associated uncertainties, and several other fitting flags are passed to the \texttt{GaussianProcessRegressor} object, as well as the kernel function to use in the fitting. Users are also able to define their own kernel function or select one (or a linear combination of multiple) from the \texttt{scikit-learn} library. As the name may imply, \texttt{GP3D} performs regression across wavelength, phase, and flux space simultaneously. Users also have the option of performing a more traditional GPR fit to a single wavelength or filter individually using \gopreaux{'s} \texttt{GP} class. For our purposes here, we will only describe the \texttt{GP3D} functionality. 

GPR constructs a final model of the input data by optimizing the hyperparameters of the input kernel. This is done by finding the hyperparameters that maximize the marginal likelihood\textemdash essentially describing how well the GPR models the data. In our fitting routine, kernel hyperparameters are optimized for a collection by individually modeling the photometry of each transient in the input collection. Final SED model surfaces for each class are produced by fixing the kernel hyperparameters to their median values and re-fitting each transient's photometry, as done in K24. To do so, best-fit SED surfaces and their uncertainties (representing 95\% confidence intervals) are first calculated for each object and smoothed using a boxcar function with a window size proportional to the number of filters fit to reduce artefacts, particularly at the boundaries of the fitting space. Next, these distributions are sampled proportionally to the number of unique photometry points used to generate them. Final SED template surfaces are calculated by median combining all the individually sampled surfaces for the objects in the processed sample. Before outputting the models, wavelengths and phases are transformed back into linear space, and flux residuals are transformed back into magnitudes relative to peak brightness in the reference blue filter.

\section{\gopreaux{} Output Models and User Actions}\label{sec:fits}

\begin{figure*}
    \vspace{-1cm}
    \centering
    \includegraphics[scale=0.66]{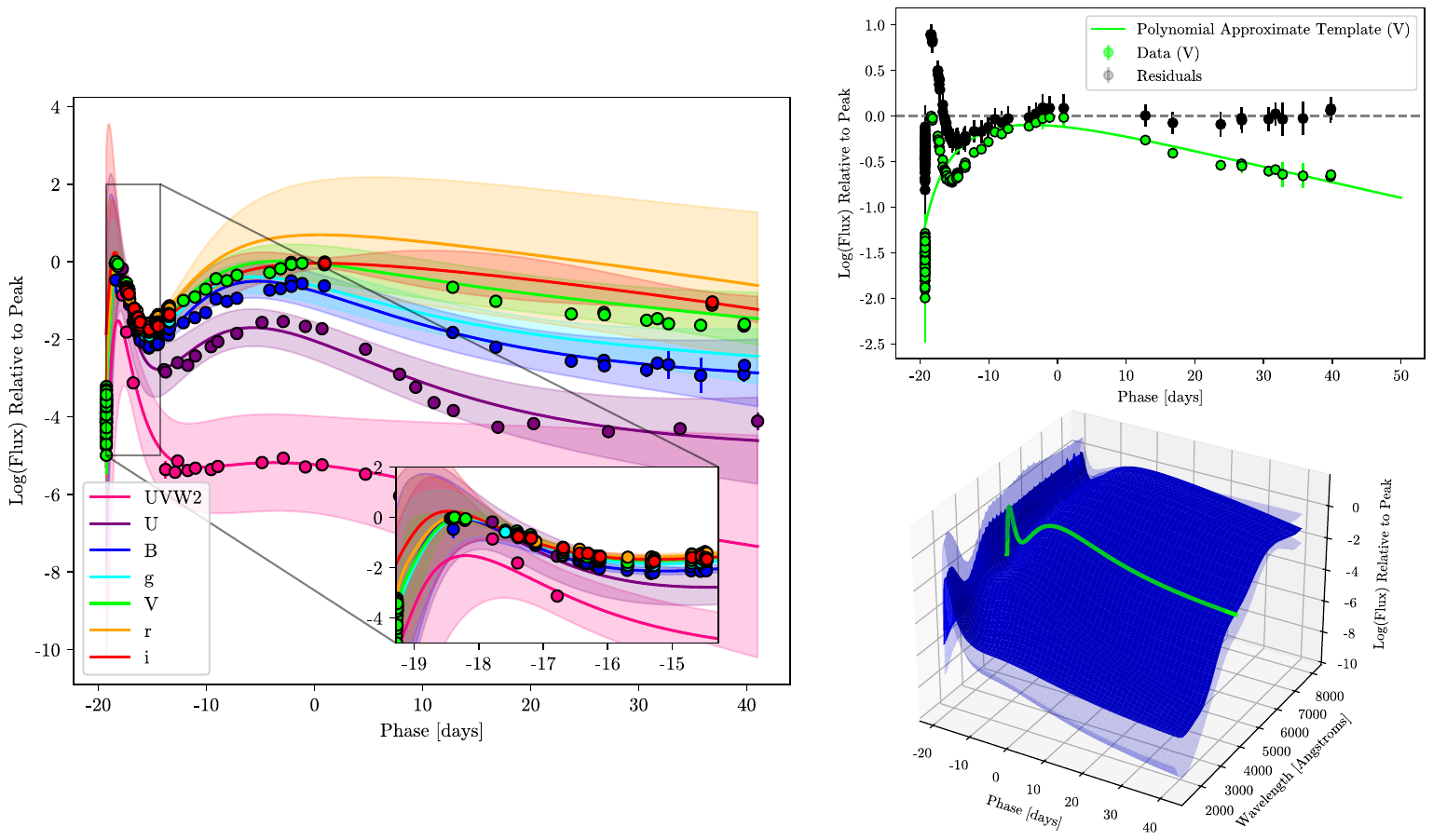}
    \vspace{-2.5cm}
    \caption{Plots demonstrating different aspects of our GPR modeling routine, from computing residuals to fitting light curve evolution and forecasting SED evolution. Left: Simultaneous GPR model fits (solid lines) to the observed UV-optical photometry of the Type IIb SN 2016gkg (colored points). The shaded region represents the 95$\%$ confidence interval for each filter. The inset focuses on the early-time evolution during the several days after explosion. The rapidly-evolving emission is well modeled across all filters. Top right: A demonstration of the processing step taken to calculate residuals for each transient's photometry. The polynomial approximation (green line), calculated from a fit to the total stripped-envelope SN photometry set, is subtracted from photometry in each band (green points, \textit{V}-band) to produce photometric residuals (black points). These residuals are the input to the Gaussian Process fitting routine. The black dashed line marks where the residuals should fall if the polynomial closely approximates the transient's light curve evolution; in this case, the late time behavior is well matched by the polynomial, but the early-time light curve excess is not. This process ensures that we are fitting our data set as residuals with a mean zero function, and that any light curve abnormalities are emphasized during our modeling. Bottom right: The final model SED surface for SN 2016gkg, from which the fits in the left panel are derived. 95$\%$ confidence intervals are again shown in the shaded region. Slices across arbitrary wavelength (or phase) gives a predicted light curve (or SED). An example showing the predicted V-band light curve is plotted in green.}
    \label{fig:fits}
\end{figure*}

An example GPR fit to observed photometry, as well as the output SED surface from that GPR across the phase and wavelength range of the fit, is shown in Figure \ref{fig:fits}. The GPR routine models and forecasts the SED evolution across the entire phase and wavelength range covered by the input photometry using the photometric residuals, and \gopreaux{} then converts these residuals back to fluxes. Each model light curve (shown in the top panel) corresponds to a ``slice" of the SED surface (shown in the bottom panel) at the effective wavelength of that filter,\footnote{Note that this is an approximation, as each photometric data point has a slightly different effective wavelength owing to our warping process described in Section \ref{sec:code}.} with an associated uncertainty corresponding to the probability distribution of the GPR at that wavelength. 

\begin{figure*}
    \centering
    \includegraphics[width=0.45\textwidth]{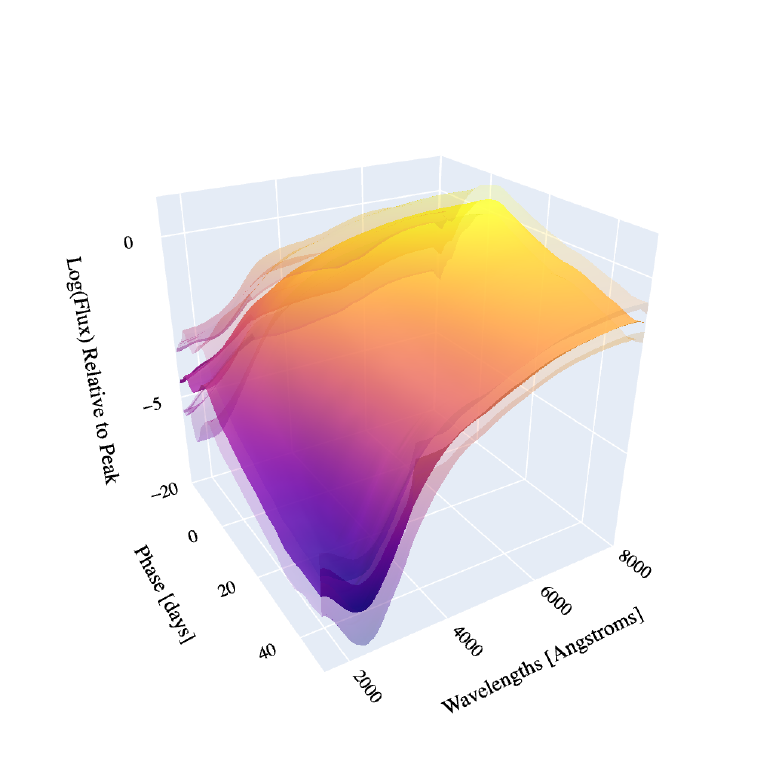}
    \includegraphics[width=0.45\textwidth]{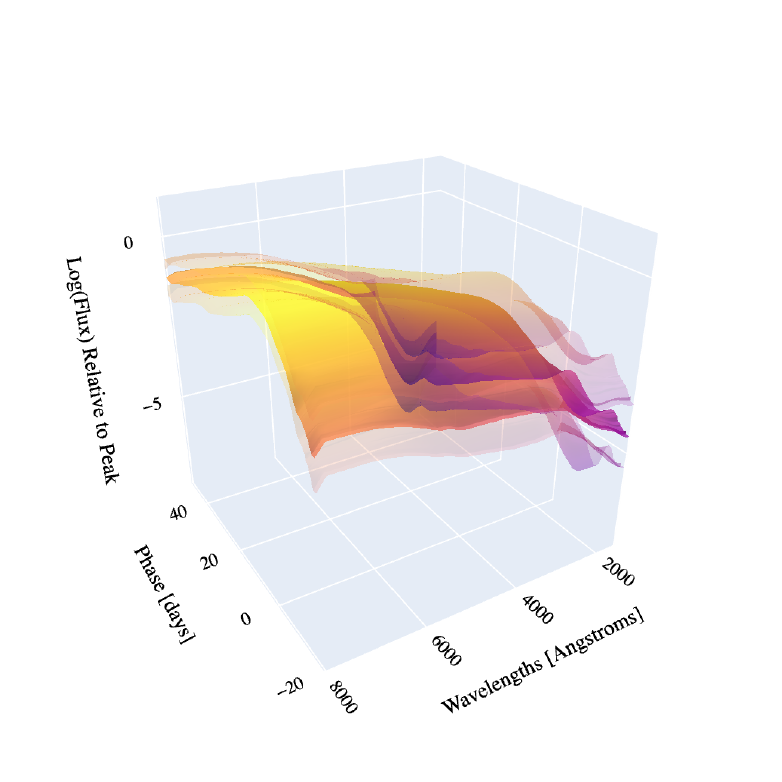}
    \includegraphics[width=0.45\textwidth]{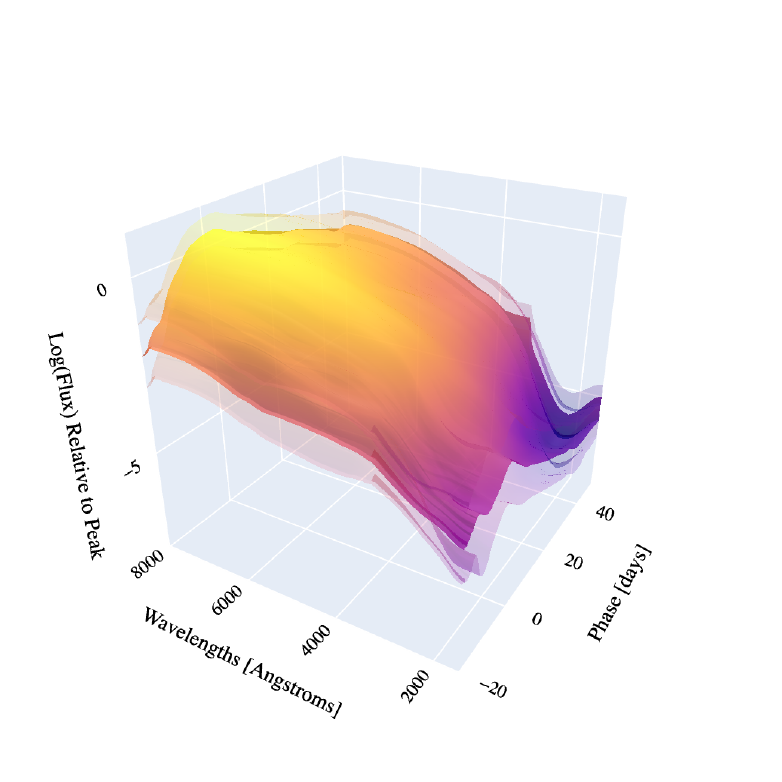}
    \caption{The final template model surface for the Type IIb supernovae, fit between -20 and 50 days relative to peak brightness across the UV and optical filters. Three different viewing angles are shown relative to the yz-plane normal: -37.5 degrees (top left), -127.5 degrees (top right), and 240 degrees (bottom). The final model is constructed by sampling and median combining the individual fits from each object in the input sample. Colors correspond to brightness relative to peak and 95\% confidence intervals are shown with greater transparency. This figure will be available as an interactive figure in the publisher version.}
    \label{fig:final_model_surface}
\end{figure*}

The SED surfaces (Figure \ref{fig:fits}, bottom) produced for each transient in the input sample are then median combined to produce a final template model surface for that group. An example of this final template model is shown in Figure \ref{fig:final_model_surface}. Each final model produced by the \texttt{GP3D.predict()} method is saved as a custom \texttt{SNModel} object. \texttt{SNModel} is a class that is used to store and initialize the Gaussian Process model. The raw output model is saved as a .fits file, and \texttt{SNModel} handles saving, loading, and initializing the model from its raw file format. Additionally, the raw file stores metadata about the output model, such as the sample used to produce it, the optimized kernel hyperparameters, and the wavelength and phase bounds of the fits, using features of .fits files such as headers and table extensions. \texttt{SNModel} also saves and loads these metadata for each final model. As part of our repository, we provide a set of uniformly-processed \texttt{SNModel} objects as .fits files for each of the main spectroscopic classes of our sample. Additional examples demonstrating the workflow of saving and loading these model objects are also provided.

Besides saving and loading the final GPR models, interested users may want to manipulate them to compare predicted light curves and SEDs to observations, sample the GP-predicted probability distribution, and generate synthetic photometry. \texttt{SNModel} has built-in routines to handle these different use cases. 

\begin{figure}
    \centering
    \includegraphics[width=1.05\linewidth]{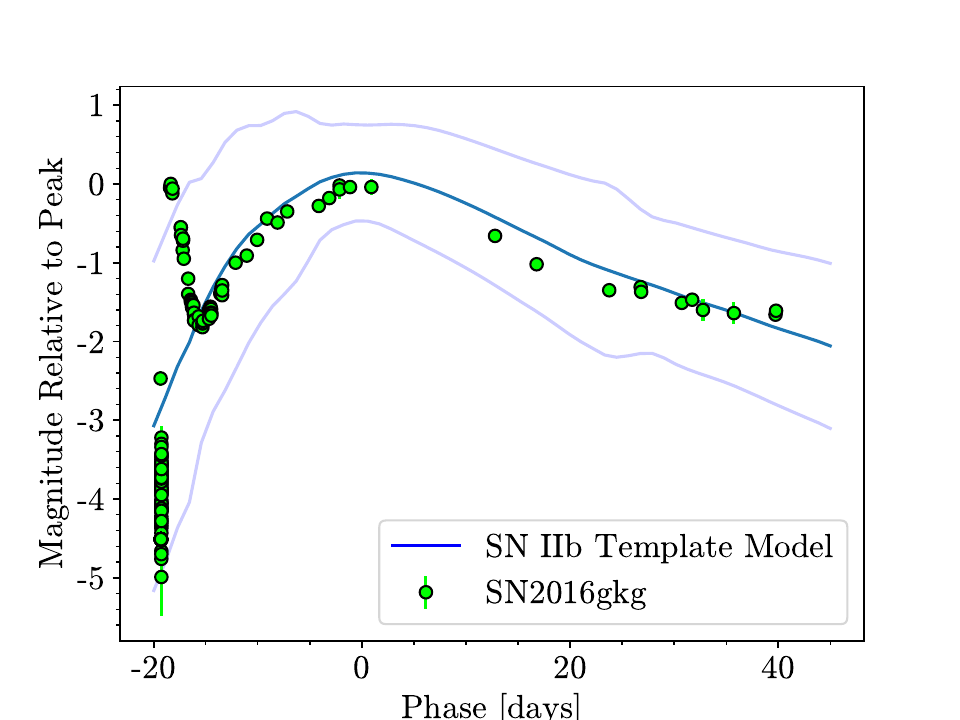}
    \caption{The observed \textit{V}-band photometry of the Type IIb SN\,2016gkg (green points) compared to the final GPR model template surface for SNe IIb at that wavelength (solid blue line). All values are plotted relative to the estimated peak brightness of the transient. 95\% confidence intervals are given by the faint blue lines. The model encompasses the observed photometry at almost all phases\textemdash the sharp initial peak is just barely outside the 95\% confidence range.}
    \label{fig:model_with_phot}
\end{figure}

Figure \ref{fig:model_with_phot} demonstrates a comparison between observed transient photometry and a model prediction for that class of transients (in this case, Type IIb SNe). This comparison is enabled by built-in methods to the \texttt{SNModel} class, allowing users to plot observations and model predictions side-by-side. The model accurately tracks the observed photometric evolution, within the confidence region, at nearly all times. This is evidence that the models are fitting the typical evolution of SNe IIb; the model has an uncertainty of around one magnitude around peak brightness, but this uncertainty increases to roughly four magnitudes at early times, reflecting both the scarcity of data at early times as well as early-time diversity in SN IIb light curve behavior relative to the secondary peak, driven by variable powering sources (e.g., \citealt{Pellegrino2023}, K24). A similar process can be used to produce a predicted SED; however, owing to the nature of our photometric standardization and the coarse resolution of our interpretation, comparing a predicted SED to an observed spectrum is more difficult. This will be explored more in the second work in this series (C. Pellegrino et al., in prep.).

\begin{figure}
    \centering
    \includegraphics[width=\linewidth]{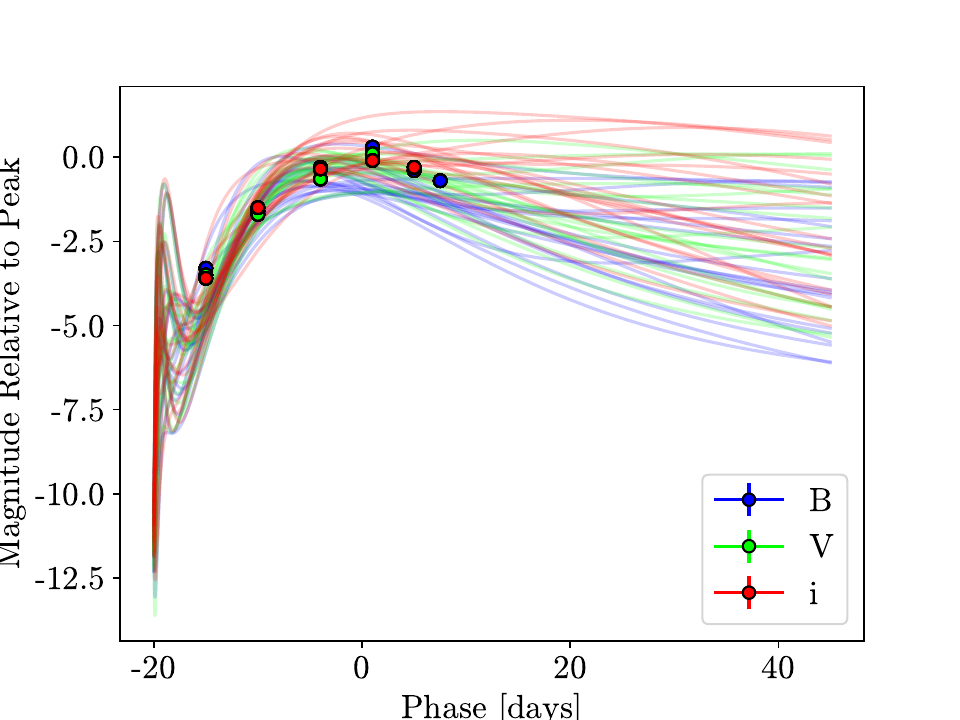}
    \caption{Predicted light curves (solid lines) generated from fits to synthetic photometry (colored points) for the final SN IIb template model. Each line represents one sample drawn from the GPR probability distribution at that filter's effective wavelength. The variable early-time light curves reflect the range of behavior for the SNe IIb at these phases; some show double-peaked light curves and others only show a smooth rise.}
    \label{fig:synthetic_lcs}
\end{figure}

\texttt{SNModel} objects can also be used to sample the probability distribution of the GPR model surfaces to produce simulated photometry points and light curves. This is shown in Figure \ref{fig:synthetic_lcs}, which plots synthetic photometry in three filters as well as 20 light curves fit to each filter, randomly sampled from the GPR probability distribution for the Type IIb SN model template. Despite the relative sparse input data, the GPR predictions match the expected photometric evolution quite well, including a potential early-time bump sometimes seen in this class of transients. This feature of \texttt{SNModel} is essential to maximizing the utility of these GPR models for upcoming all-sky surveys by producing synthetic photometry for different classes of transients, testing survey cadences and identification strategies to photometrically classify transients with limited data.

Finally, \gopreaux{} is designed to be flexible enough so that users can add new transients and fit their photometry seamlessly. To accomplish this, the codebase has built-in functionality to automatically process data for new transients from major sources described here, such as Swift, all-sky surveys, as well as any photometry in the same \texttt{.csv} format as provided by the Open SN Catalog. Users can perform all of the pre-processing functionality described in Section \ref{sec:code} programmatically, add new transients to the CAAT file, and utilize the \texttt{GP3D} fitting routines for those objects with no discernible difference. For example, data dictionaries can be passed to the \gopreaux{} \texttt{SN} class, initialized with information about the transient classification, redshift, and coordinates. This data can then be fit for the time of peak brightness, and the shifted data can be saved to the \texttt{CAAT} archive along with the fitted parameters. Examples demonstrating this workflow will be available as part of the code repository. Additionally, we provide all our processed data for our sample in this repository to allow users to reproduce our fits, or produce new fits, using these already processed transients. The results of the fitting with our included data will be discussed in the next paper in C. Pellegrino et al. 2026, in prep.

\section{Discussion and Conclusions}\label{sec:discussion}

\subsection{Model Validation}

\begin{figure}
    \centering
    \includegraphics[width=1.05\linewidth]{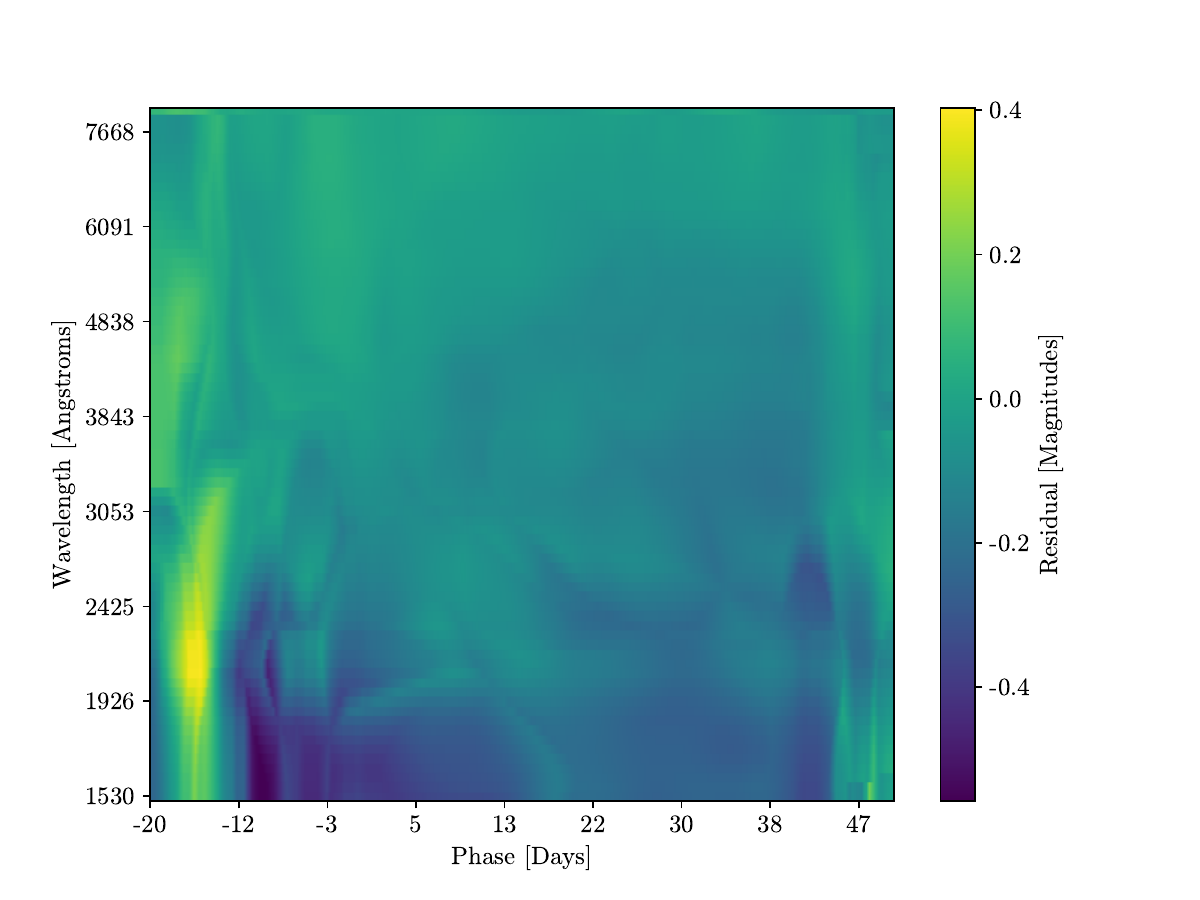}
    \caption{Heatmap showing the median residuals between five train-test splits of the SN IIb sample. The magnitude of the residuals is largest in the bluest wavelengths, particularly at the phase boundaries, owing to a scarcity of data in this region of parameter space.}
    \label{fig:traintest}
\end{figure}

To estimate the effect different collections of input objects have on the outputted model, we run a series of train/test splits for several of the classes in our sample. For a given class, we randomly select 80$\%$ of its objects and perform our GPR fitting. We repeat this process five times, each time selecting a new random sample from the total number of objects in that class. We then difference the output SED surfaces with each other to obtain model residuals, which we take to be the average deviation at each wavelength and phase step caused by differences in the inputted transient collections.

An example of the results of this process is shown in Figure \ref{fig:traintest}. For most classes of objects, these residuals are no more than $\pm$ 0.2 magnitudes across our phase and wavelength ranges\textemdash well within the $95\%$ confidence intervals of our GPR routine. We notice larger residuals at bluer wavelengths, particularly within the \textit{Swift} \textit{UVW2}-band. This is likely due to a lack of data in these filters, particularly at later phases, as well as the complex photometric behavior caused by the rapidly-cooling transient emission. Likewise, we notice some deviations at late phases which is likely caused by different amounts of extrapolation, depending on the number of inputted objects with well-sampled late-phase light curves.

In general, we find that the classes with larger samples of objects\textemdash SNe II and stripped-envelope SNe in particular\textemdash have smaller train/test split residuals than classes with smaller samples, such as FBOTs. This is expected, as our process of randomly including or excluding certain objects will have a greater effect on the output model when there are relatively few objects with well-sampled multi-band light curves. This process demonstrates that our modeling routine is robust and reproducible, and that building larger samples of transients across spectroscopic classes will lead to reduced statistical uncertainties in the future.

\subsection{Sample and Methodology Limitations}

Our sample selection criteria and modeling routines lead to several limitations in the output transient models. One such factor is the necessity to model magnitudes relative to peak brightness in a certain filter. This is done to standardize all transient light curves across classes, by comparing to the light curve peak in the approximate rest-frame \textit{V}-band. However, this leads to certain drawbacks. First is that we must exclude transients without sufficient coverage around the light curve peak in this wavelength region, thereby limiting our final sample of transients we fit. This also makes comparisons between transients' absolute brightnesses difficult; we do not record or incorporate any information about absolute magnitudes or luminosities in our fitting routine. Therefore, all comparisons between transients' light curve shapes, rise times, and decline times are done relative to the time of peak in the optical regime.

Likewise, our reliance on the knowledge of the time and brightness of the light curve peak makes modeling and comparisons between ``early" light curve peaks difficult. For example, comparing only a handful of light curve points on the rise to peak with an output model is difficult, as there is no knowledge of the transient's peak information until it is observed. In other words, without peak information, there is both an uncertainty in relative magnitude and relative phase, both compared to peak. 

Yet another limitation in our results arises from systematic biases in our photometry. Perhaps the most obvious example is host galaxy dust extinction. While \gopreaux{} corrects for Milky Way extinction, no corrections are made for the transients' host galaxy extinction. Such corrections are difficult to do systematically, particularly without high-resolution spectra or prior knowledge of a transient class's intrinsic color evolution. Incorporating either of these would severely limit our sample or bias our results by introducing assumptions about the physics of the objects. However, by not adjusting for host galaxy extinction, our models will have a larger spread in the measured relative brightness of all transients, particularly in bluer filters, where the effect of dust attenuation is greater \citep[e.g.,][]{Cardelli1989}.

Finally, by limiting our sample to those transients with \textit{Swift} observations, we are implicitly biasing our sample toward low-redshift objects, which are more likely to have \textit{Swift} data requested. Because of this, we may not be capturing redshift dependence in our data or sample of objects. While the former effect is likely to be negligible for classes of core-collapse SNe, which have larger intrinsic diversities in their light curve evolution, there is evidence for some redshift dependence among Type Ia SNe \citep[e.g.,][]{Howell2007,Nicolas2021}, which we likely do not fully capture. The latter point\textemdash that the intrinsic rates, and therefore the sample sizes, of transients evolve with redshift \citep[e.g.,][]{Strolger2015}\textemdash may also bias our sample. For example, we have lower sample sizes of transients that are rarer at low redshifts, including SLSN-II. This may affect our modeling efforts of these classes by under-predicting or over-predicting their inherent photometric diversity. 

These limitations can be summarized as follows: \gopreaux{} is designed to model \textit{observed} photometry of transients, with as few implicit assumptions made about the underlying physics, or populations these transients are drawn from, as possible. This methodology choice was a conscious effort, meant to complement the nature of GPR by not introducing any unnecessary assumptions about the underlying physics of the objects we model. This reliance on observed data only is best suited to enable classification and regression of future transients discovered by current and future time-domain surveys, which will have limited information beyond apparent magnitudes in several filters. \gopreaux{} was designed with these assumptions and limitations in mind.

\subsection{Concluding Remarks}

We have introduced \gopreaux{}, a modular, open-source software package that models the SED evolution of extragalactic transients simultaneously in phase and wavelength using Gaussian Process Regression. As a result of this effort, this work provides the following:
\begin{itemize}
    \item a uniformly created set of three-dimensional SED ``surfaces'' which cover a large range of wavelengths and phases for all major transient classes;
    \item uniformly processed ultraviolet to infrared photometry for over 1,300 extragalactic transients, encompassing over 140,000 observations, which can be used for future modeling and analysis; and
    \item modularized and open-source code to enable modeling for an arbitrary collection of transients, independent of their spectroscopic classifications or underlying physics.
\end{itemize}

By forecasting transients' rest-frame emission from the ultraviolet to the infrared, \gopreaux{} enables photometric classification and parameter inference for the large populations of transients that will be observed by the Rubin Observatory and \textit{Roman} Space Telescope over the coming decade. To enable this effort, we have assembled one of the largest samples of extragalactic transient photometry. Our database, a.k.a. the Catalog of Archival Astronomical Transients, includes almost 1,300 objects and a total of 146,000 photometry points, all uniformly processed and made available to the community as part of our code repository. 

The overall goal of this paper is to present and analyze SED template models for the different transient classifications described herein. \gopreaux{} and the templates we present here will enable a variety of studies. We are developing an in-depth comparison of our most up-to-date template models with existing models in the literature, with a particular focus on core-collapse SNe, which we reserve for an upcoming publication. Additionally, our modeling methodology can be used to generate synthetic datasets to simulate transients discovered at high redshifts, and to inform photometric classifiers to better identify transients of interest in time-domain alert streams. The models can also be used to perform population-level analysis of the physical mechanisms powering transient emission, including physical parameter inference. Finally, bolometric corrections as a function of transient type, subtype, or a number of other features can be calculated to better understand their rest-frame behavior.

The modularity of this codebase allows for new transients (or transient classes) to be added in the future. While the data and the classifications we have set up are useful for our intended analyses, new users are encouraged to modify \gopreaux{} to fit their own specific research needs. Through the flexibility of this repository, combined with the non-parametric, data-driven nature of Gaussian Processes, we hope to enable new advances in population-level modeling across all areas of time-domain astronomy.

\begin{acknowledgments}
This work is supported by ADAP program grant No. 80NSSC24K0180 and made possible in part by NASA Research Announcement NNH23ZDA001N.
M.M. and the METAL group at UVA acknowledge support in part from ADAP program grant No. 80NSSC22K0486, from the NSF grant AST-2206657 and from the National Science Foundation under Cooperative Agreement 2421782 and the Simons Foundation grant MPS-AI-00010515 awarded to the NSF-Simons AI Institute for Cosmic Origins — CosmicAI, https://www.cosmicai.org/.
This research has made use of the Astrophysics Data System, funded by NASA under Cooperative Agreement 80NSSC25M7105. S.K. was funded by an LSST-DA Catalyst Fellowship through Grant 62192 from the John Templeton Foundation to LSST Discovery Alliance when contributing to this project.

\end{acknowledgments}

\bibliographystyle{aasjournalv7}
\bibliography{references}

\software{
\texttt{scikit-learn} \citep{Pedregosa2011}, \texttt{matplotlib} \citep{Hunter:2007}, \texttt{numpy} \citep{Harris2020}, \texttt{pandas} \citep{reback2020pandas}, \texttt{astropy} \citep{astropy:2013,astropy:2018,astropy:2022}, \texttt{scipy} \citep{2020SciPy-NMeth}
}

\appendix

\restartappendixnumbering

\section{Supplemental Figures}

\begin{figure}
    \centering
    \includegraphics[width=0.95\linewidth]{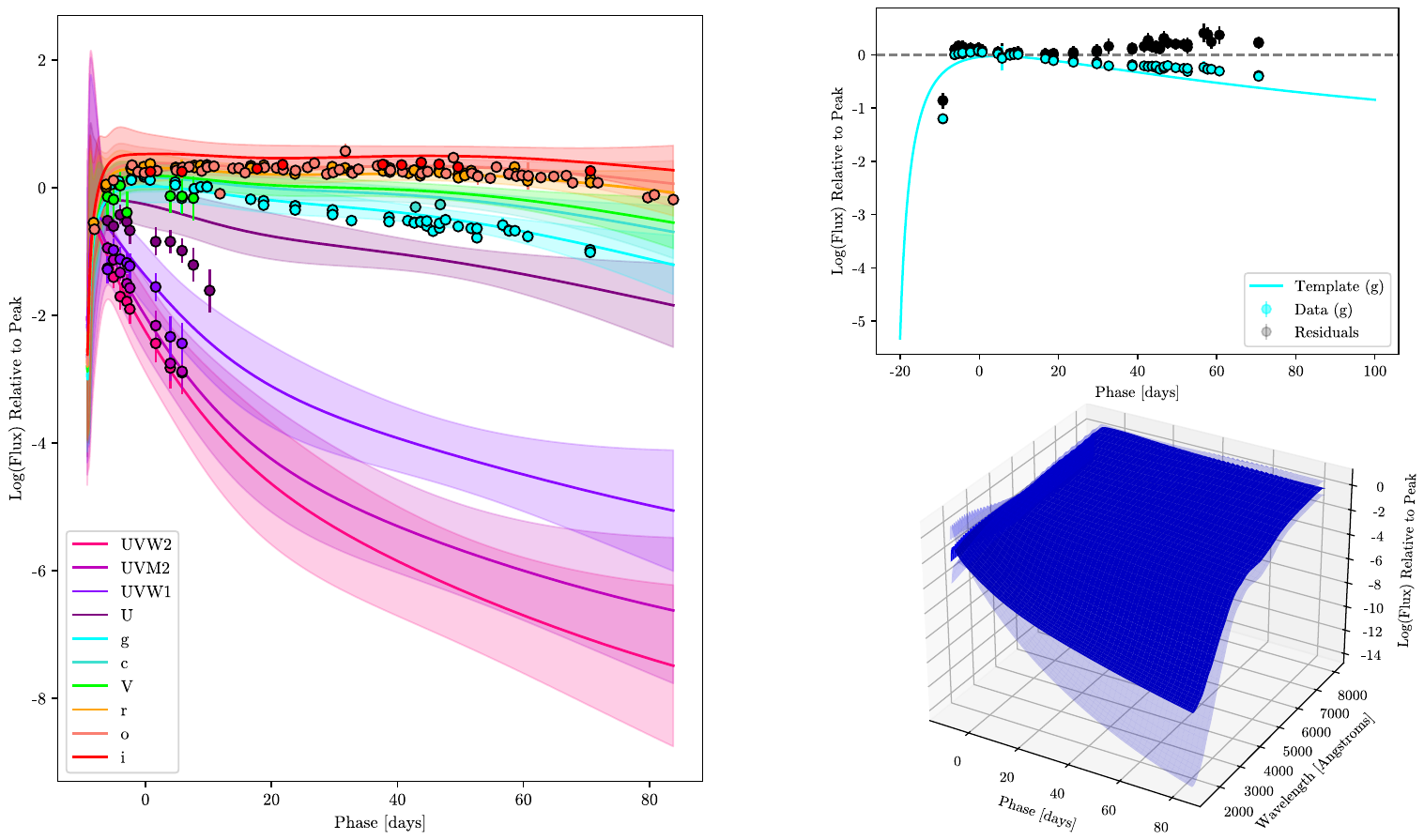}
    \caption{Same as Figure \ref{fig:fits} but for the Type II SN 2021yyg.}
    \label{fig:sniifits}
\end{figure}

\begin{figure}
    \centering
    \includegraphics[width=0.95\linewidth]{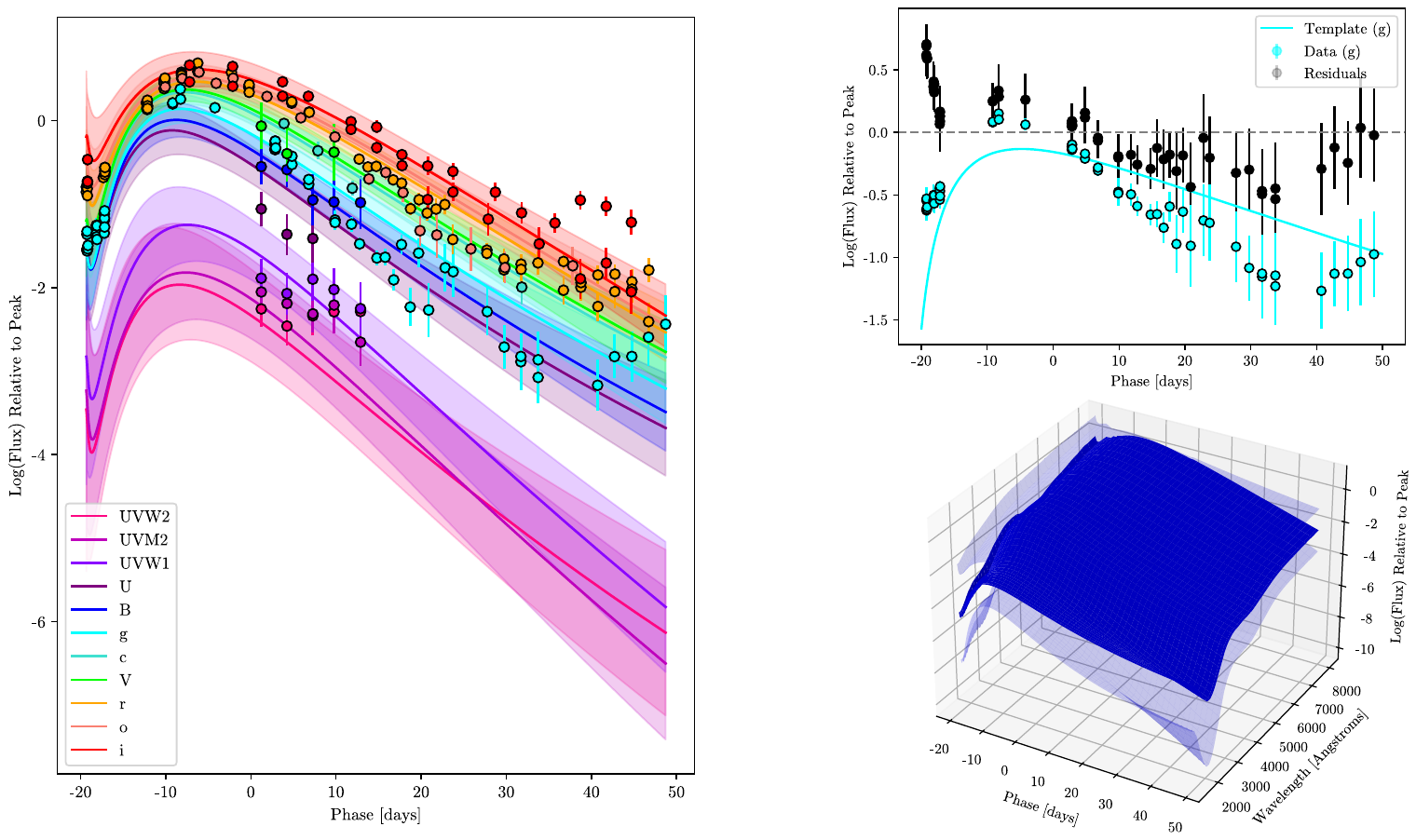}
    \caption{Same as Figure \ref{fig:fits} but for the Type Ib SN 2022hgk.}
    \label{fig:snibfits}
\end{figure}

\begin{figure}
    \centering
    \includegraphics[width=0.95\linewidth]{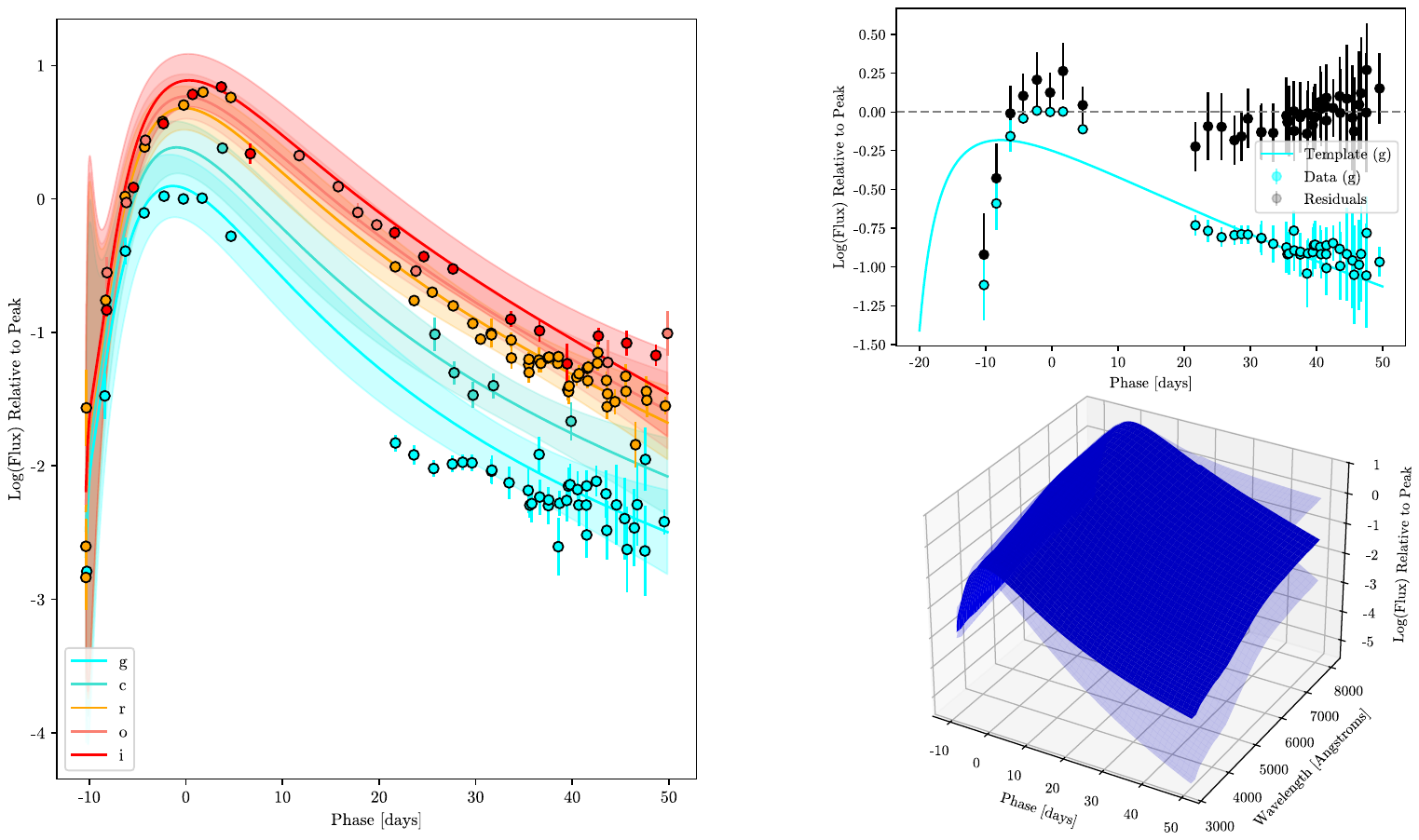}
    \caption{Same as Figure \ref{fig:fits} but for the Type Ic SN 2021do.}
    \label{fig:snicfits}
\end{figure}

\begin{figure}
    \centering
    \includegraphics[width=0.95\linewidth]{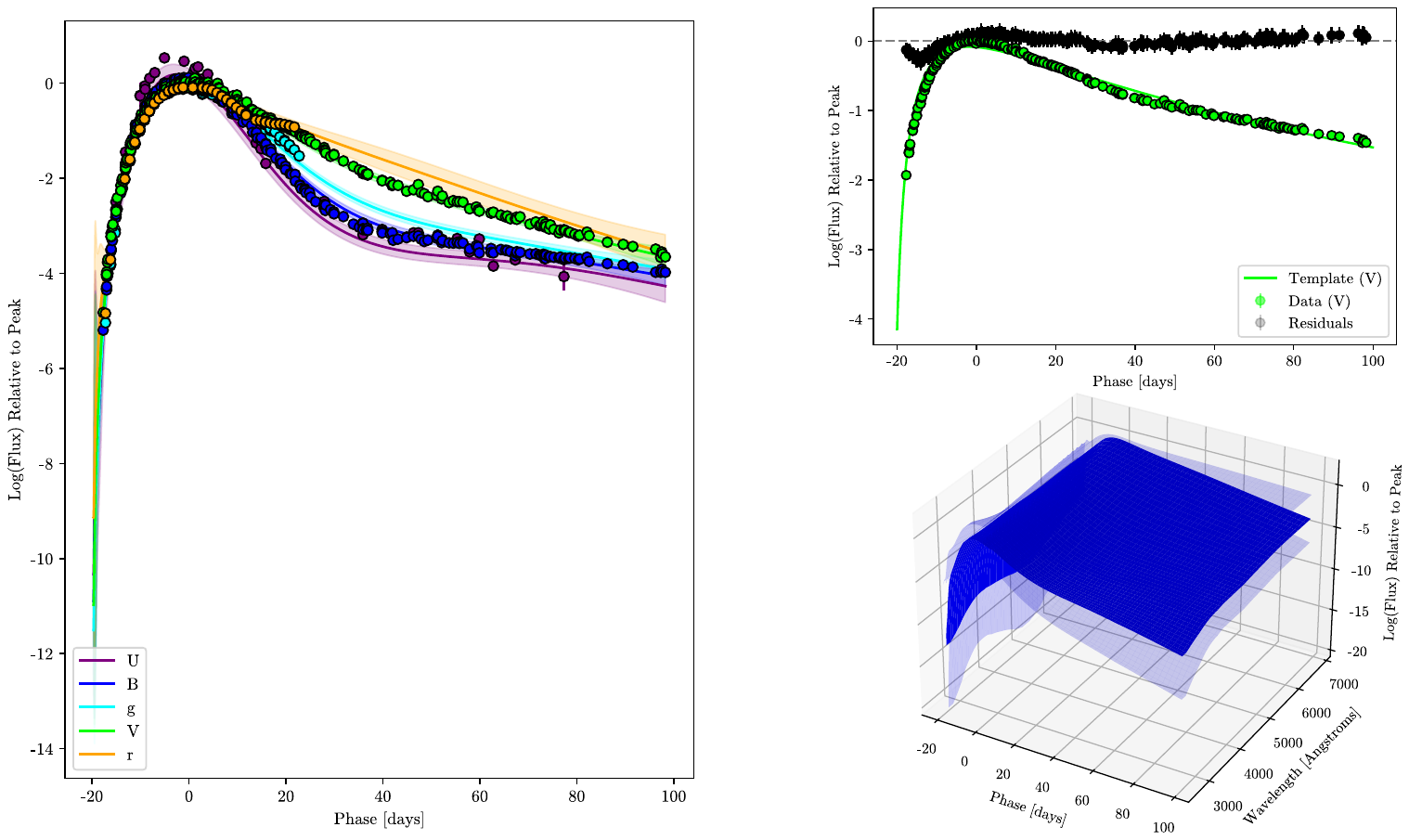}
    \caption{Same as Figure \ref{fig:fits} but for the Type Ia SN 2011fe.}
    \label{fig:sniafits}
\end{figure}

\begin{figure}
    \centering
    \includegraphics[width=0.95\linewidth]{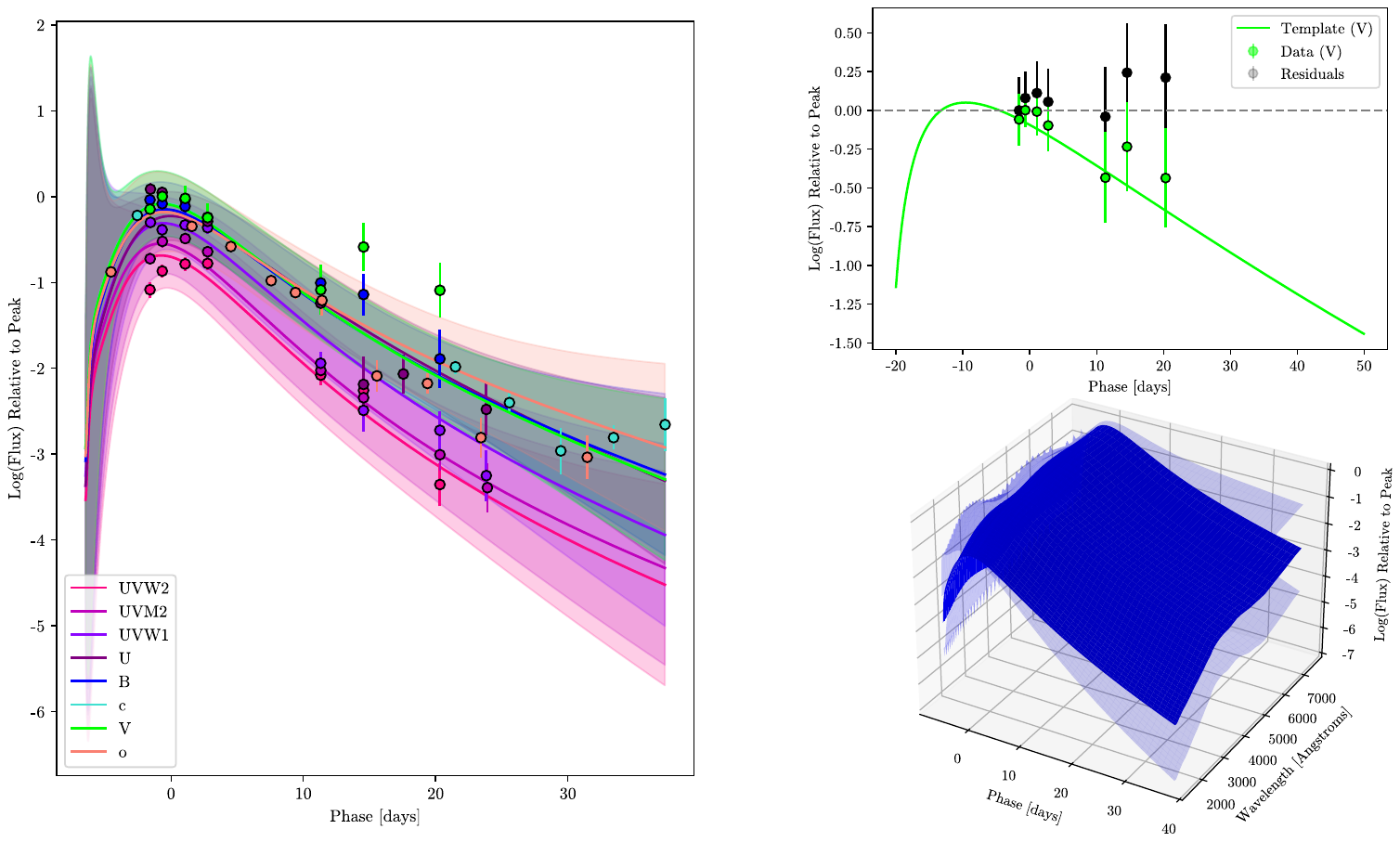}
    \caption{Same as Figure \ref{fig:fits} but for the Type Ibn SN 2019kbj.}
    \label{fig:snibnfits}
\end{figure}

\begin{figure}
    \centering
    \includegraphics[width=0.95\linewidth]{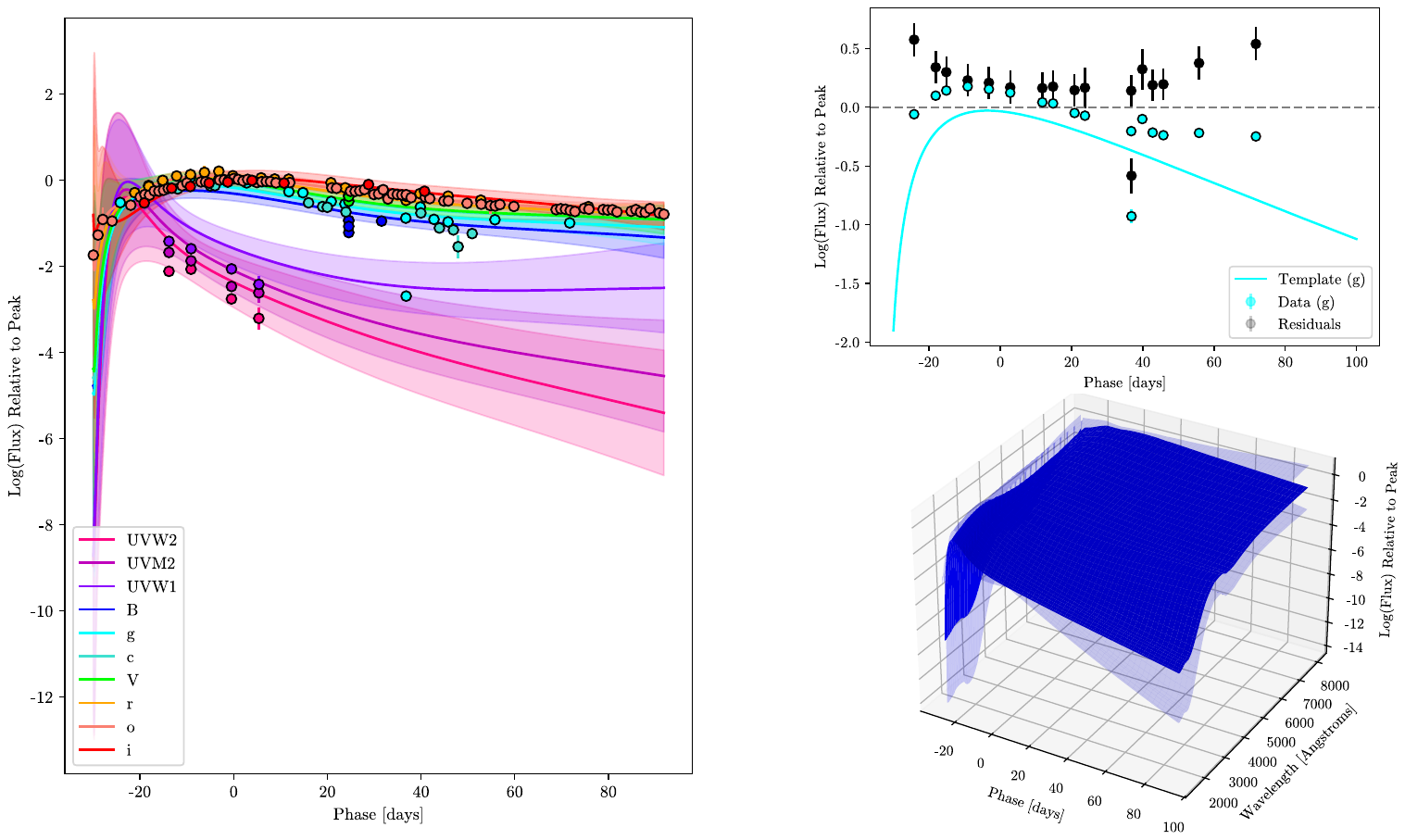}
    \caption{Same as Figure \ref{fig:fits} but for the Type IIn SN 2018bwr.}
    \label{fig:sniinfits}
\end{figure}

\begin{figure}
    \centering
    \includegraphics[width=0.95\linewidth]{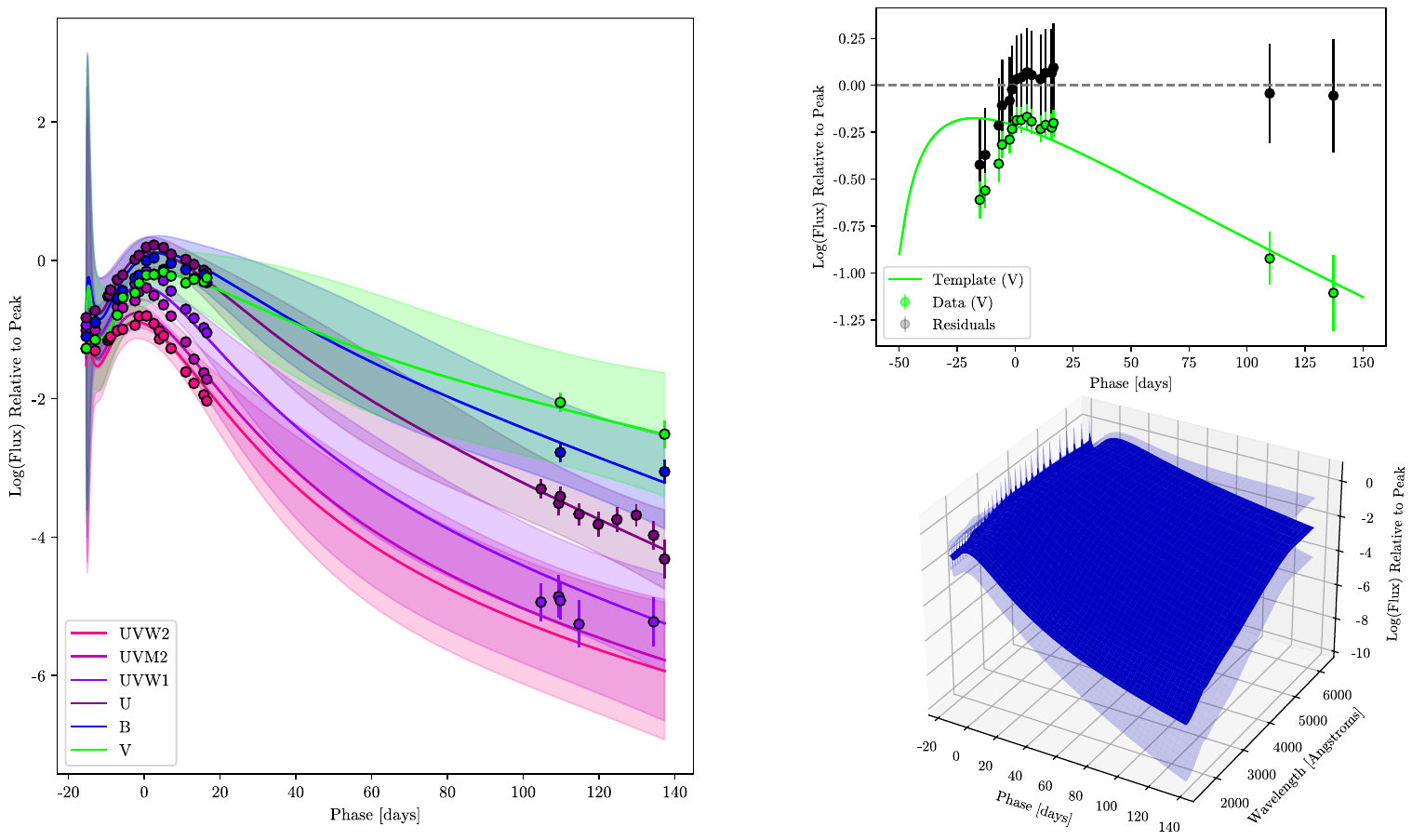}
    \caption{Same as Figure \ref{fig:fits} but for the Type I SLSN 2017egm.}
    \label{fig:slsnifits}
\end{figure}

\begin{figure}
    \centering
    \includegraphics[width=0.95\linewidth]{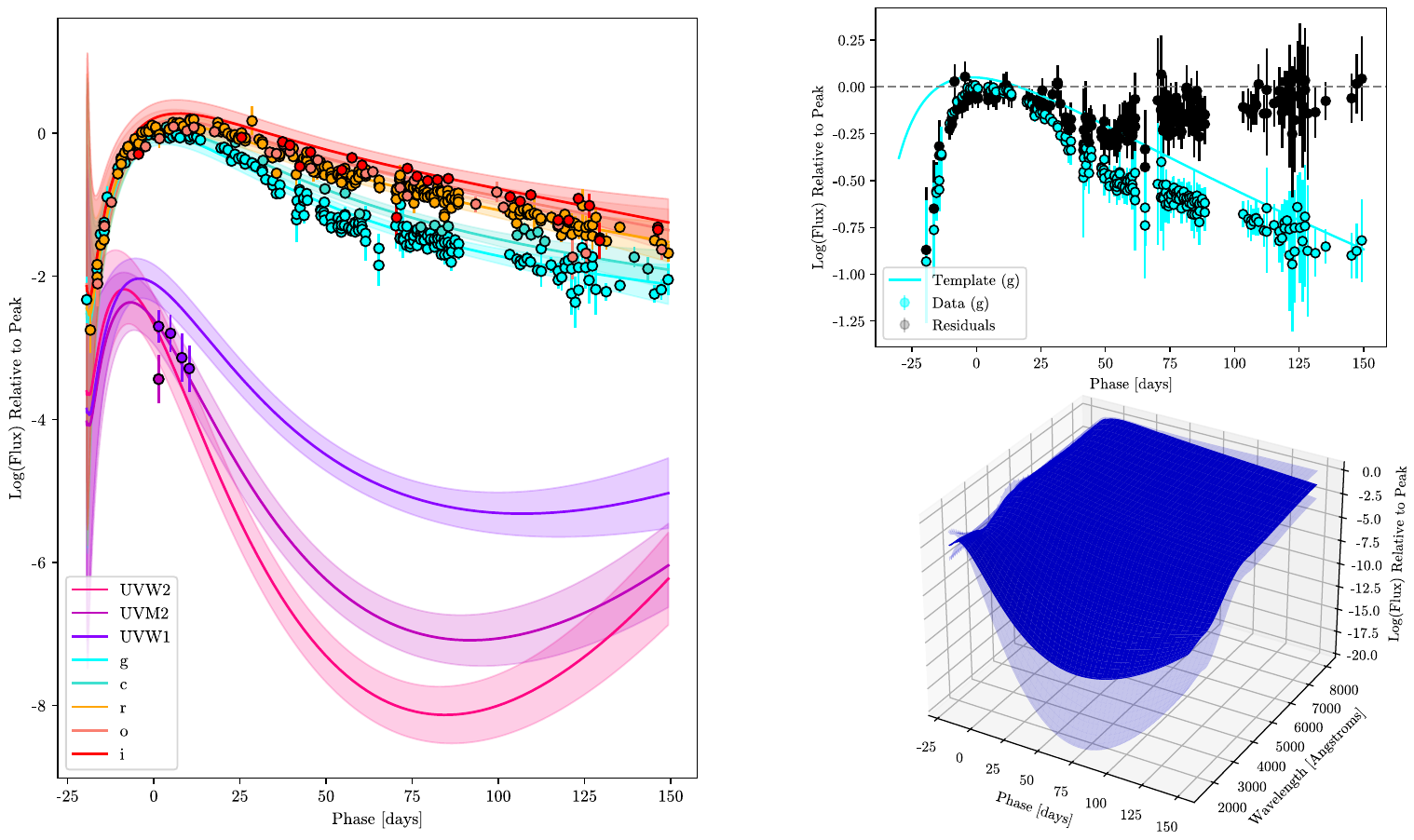}
    \caption{Same as Figure \ref{fig:fits} but for the Type II SLSN 2020vfu.}
    \label{fig:slsniifits}
\end{figure}

\begin{figure}
    \centering
    \includegraphics[width=0.95\linewidth]{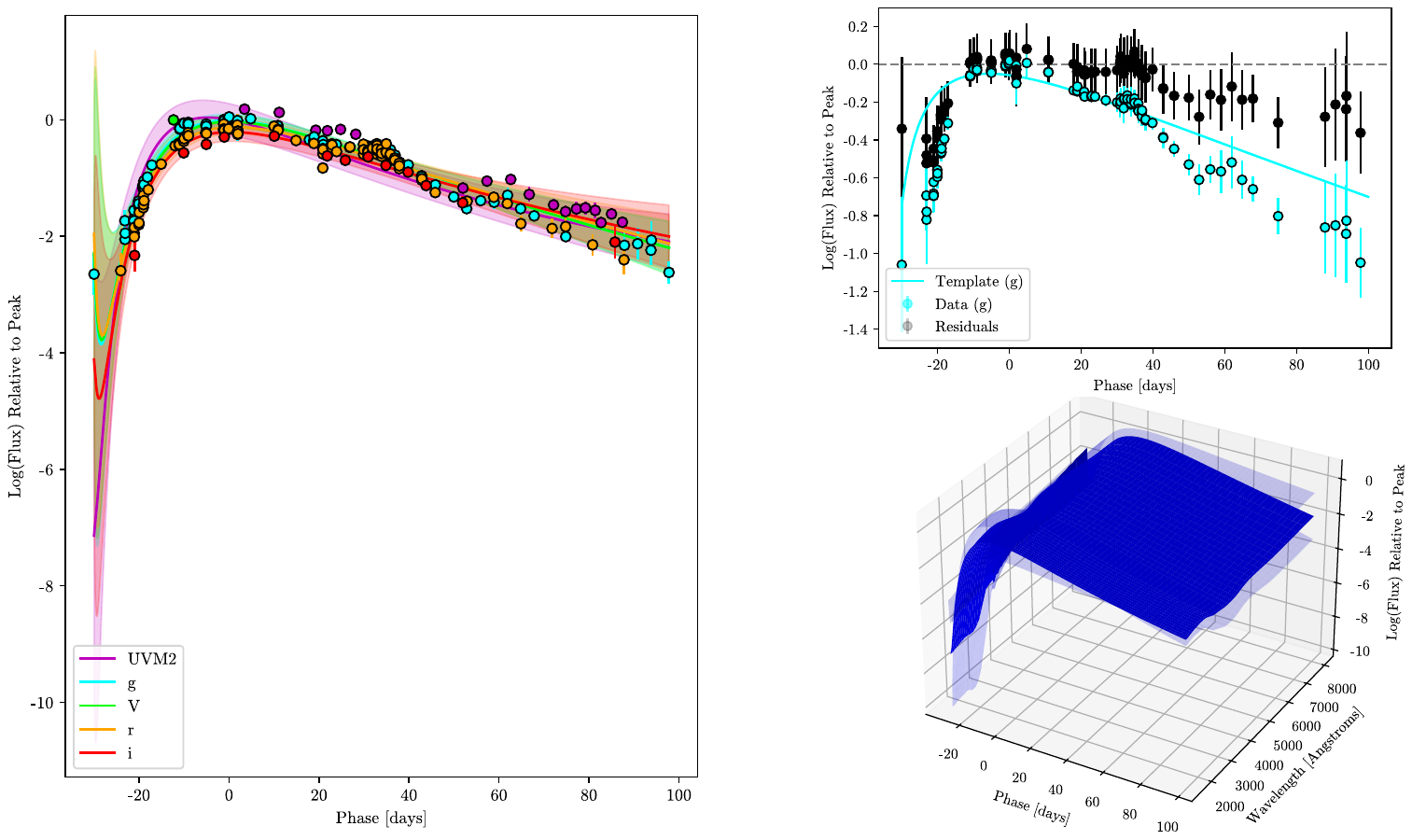}
    \caption{Same as Figure \ref{fig:fits} but for the TDE AT 2019ehz.}
    \label{fig:tdefits}
\end{figure}

In this Section we provide example GPR fits to different types of transients in our sample (Figures \ref{fig:sniifits} - \ref{fig:tdefits}). Note that each fit uses the same kernel as the Type IIb model, and therefore show large uncertainties at early times. These plots are meant to merely show examples of fits from different classes, not perform vetting and comparisons of our models. We leave a thorough description of each template model and their optimized kernel hyperparameters to a future work (C. Pellegrino et al., in prep.).

\section{Supplemental Tables}

\clearpage

\footnotesize
\begin{longrotatetable}
\begin{deluxetable*}{ccccccccccccc}
    \tabletypesize{\scriptsize}
    \movetabledown=8mm
    
    \tablecaption{Sample Description}
    \label{tbl:sampleproperties}
    \tablehead{
        \colhead{Name} & \colhead{Type} & \colhead{Subtype} & \colhead{Redshift} & \colhead{Peak} & \colhead{Peak} & \colhead{Peak} & \colhead{\# Detections} & \colhead{Filters with} & \colhead{Phase of 1st} & \colhead{Phase of Last} & \colhead{Data} & \colhead{References} \\
        & & & & \colhead{MJD} & \colhead{Mag} & \colhead{Filter} & & \colhead{Detections} & \colhead{Detection} & \colhead{Detection} & \colhead{Sources} & \\
    }

    \startdata
    SN2022qzr & SESNe & SNIIb & 0.019 & 59820.6 & 17.41 & g & 141 & g,c,r,o & -19.8 & 311.8 & ATLAS, Swift, ZTF & \\ 
    \hline
    SN2022ngb & SESNe & SNIIb & 0.009 & 59772.3 & 16.74 & g & 270 & B,g,c,V,r,o,i & -1494.8 & 367.9 & ATLAS, Swift, ZTF & \\ 
    \hline
    SN2022hnt & SESNe & SNIIb & 0.0192 & 59707.2 & 17.91 & o & 37 & \shortstack{UVW2,UVM2,UVW1 \\ U,B,c,V,o} & -23.8 & 381.1 & ATLAS, Swift & \\ 
    \hline
    SN2022eji & SESNe & SNIIb & 0.017 & 59662.1 & 17.16 & c & 222 & \shortstack{UVW2,UVM2,UVW1 \\ U,B,g,c,V,r,G,o,i} & -1231.7 & 393.1 & \shortstack{ASAS-SN, ATLAS \\ OSC, Swift, ZTF} & \\ 
    \hline
    SN2022crv & SESNe & SNIIb & 0.0080 & 59654.2 & 15.45 & o & 75 & UVW2,UVW1,c,o & -27.7 & 317.8 & ATLAS, Swift & \\ 
    \hline
    SN2021ybc & SESNe & SNIIb & 0.0292 & 59474.3 & 18.56 & g & 69 & U,B,g,c,r,o & -10.9 & 424.1 & ATLAS, Swift, ZTF & \\ 
    \hline
    SN2021uqw & SESNe & SNIIb & 0.0280 & 59432.5 & 17.98 & r & 42 & \shortstack{UVW2,UVM2,UVW1 \\ U,B,g,r,i} & -1142.1 & 25.8 & \shortstack{ASAS-SN, Swift \\ ZTF} & \\ 
    \hline
    SN2021sjt & SESNe & SNIIb & 0.0047 & 59409.1 & 18.26 & c & 116 & UVW1,U,B,g,c,r,o & -2292.2 & 135.1 & ATLAS, Swift, ZTF & \\ 
    \hline
    \enddata

    \tablecomments{The table is available in its entirety in the online format. A portion is shown here for clarity, along with all references.}
    
    [1]: \citet{2018Natur.554..497B}; [2]: \citet{2014ApSS.354...89B}; [3]: \citet{2016ATel.9526....1T}; [4]: \citet{2016ATel.9521....1N}; [5]: \citet{2016ATel.9536....1K}; [6]: \citet{Arcavi2017}; [7]: \citet{2016ATel.8944....1C}; [8]: \citet{2014MNRAS.439.1807B}; [9]: \citet{2013MNRAS.433....2S}; [10]: \citet{2016ApJ...833..231T}; [11]: \citet{2015AA...580A.142E}; [12]: \citet{2013ApJ...778L..19H}; [13]: \citet{2019ApJS..241...38S}; [14]: \citet{2014ApJS..213...19B}; [15]: \citet{2018AA...609A.134S}; [16]: \citet{2009ApJ...696..870D}; [17]: \citet{2011MNRAS.413.2583S}; [18]: \citet{2011ApJ...741...97D}; [19]: \citet{2005PASP..117..132R}; [20]: \citet{2018ApJ...860...90V}; [21]: \citet{2011ApJ...728...14P}; [22]: \citet{2010AA...512A..70Y}; [23]: \citet{2019MNRAS.490.2799D}; [24]: \citet{2012MNRAS.425.1789S}; [25]: \citet{2018arXiv180100340T}; [26]: \citet{2016ATel.8935....1M}; [27]: \citet{2015ApJ...807...59H}; [28]: \citet{2016MNRAS.461.2003Y}; [29]: \citet{2017ApJ...834..118M}; [30]: \citet{2015MNRAS.450.2373B}; [31]: \citet{2017ApJS..233....6H}; [32]: \citet{2013AA...555A.142I}; [33]: \citet{2012MNRAS.422.1122I}; [34]: \citet{2014ApJ...786...67A}; [35]: \citet{2010ApJ...715..541A}; [36]: \citet{2011MNRAS.417..261I}; [37]: \citet{2013AA...549A..79S}; [38]: \citet{2012ApJ...753..109G}; [39]: \citet{2009MNRAS.394.2266P}; [40]: \citet{2018AA...609A.134S}; [41]: \citet{2014MNRAS.442..844F}; [42]: \citet{2008ApJ...675..644D}; [43]: \citet{2018MNRAS.475.3959H}; [44]: \citet{2016ApJ...820...33R}; [45]: \citet{2016MNRAS.459.3939V}; [46]: \citet{2013ATel.5455....1C}; [47]: \citet{2014MNRAS.438..368T}; [48]: \citet{2011AA...527A..61S}; [49]: \citet{2011ApJ...732..109M}; [50]: \citet{2006AJ....131.2245Z}; [51]: \citet{2020NatAs...4..893N}; [52]: \citet{2016ATel.8790....1C}; [53]: \citet{2014ApJ...789..104O}; [54]: \citet{2013AJ....146....2F}; [55]: \citet{2011ApJ...730...34S}; [56]: \citet{2014ApJ...797..118F}; [57]: \citet{2012AJ....144..131Z}; [58]: \citet{2011AJ....142...45A}; [59]: \citet{2010arXiv1007.0011P}; [60]: \citet{2009MNRAS.398.1041B}; [61]: \citet{2011ApJ...741....7F}; [62]: \citet{2012ApJ...756..173S}; [63]: \citet{2010ApJ...725.1768F}; [64]: \citet{2012MNRAS.426.1905S}; [65]: \citet{2020MNRAS.497..318L}; [66]: \citet{2018AA...620A..67A}; [67]: \citet{2016ApJ...826...39N}; [68]: \citet{2016ApJ...828L..18N}; [69]: \citet{2018ApJ...866L..24N}; [70]: \citet{2019MNRAS.490.3882S}; [71]: \citet{2018MNRAS.475..193F}; [72]: \citet{2016ATel.8628....1F}; [73]: \citet{2016ATel.9049....1P}; [74]: \citet{2016ATel.8885....1B}; [75]: \citet{2018PASP..130f4101V}; [76]: \citet{2014CoSka..44...67T}; [77]: \citet{2014AJ....148....1Z}; [78]: \citet{2017MNRAS.466.3442J}; [79]: \citet{2014ApJ...784L..12G}; [80]: \citet{2014MNRAS.443.2887F}; [81]: \citet{2013ATel.5637....1Z}; [82]: \citet{2017MNRAS.472.3437G}; [83]: \citet{2015AA...579A..40S}; [84]: \citet{2018ApJ...869...56B}; [85]: \citet{2018ApJ...859...24C}; [86]: \citet{2013NewA...20...30M}; [87]: \citet{2016ApJ...819...31G}; [88]: \citet{2014MNRAS.444.3258M}; [89]: \citet{2015MNRAS.446.3895F}; [90]: \citet{2014ApJ...784..105W}; [91]: \citet{2018AJ....155..201W}; [92]: \citet{2012JAVSO..40..872R}; [93]: \citet{2011Natur.480..344N}; [94]: \citet{2013ApJ...767..119M}; [95]: \citet{2013CoSka..43...94T}; [96]: \citet{2015ApJS..220....9F}; [97]: \citet{2016AA...590A...5G}; [98]: \citet{2012ApJS..200...12H}; [99]: \citet{2012ApJ...749...18B}; [100]: \citet{2010AJ....139..519C}; [101]: \citet{2010ApJS..190..418G}; [102]: \citet{2009ApJ...700..331H}; [103]: \citet{2011PhDT........35K}; [104]: \citet{2018ApJ...852..100M}; [105]: \citet{2016ATel.8907....1M}; [106]: \citet{2016ATel.8929....1V}; [107]: \citet{2018MNRAS.480.1445D}; [108]: \citet{2016PASJ...68...68Y}; [109]: \citet{2014MNRAS.443.1663C}; [110]: \citet{2011MNRAS.410..585S}; [111]: \citet{2015AA...573A...2S}; [112]: \citet{2013ApJ...767...57F}; [113]: \citet{2009AJ....138..376F}; [114]: \citet{2010AJ....140.1321F}; [115]: \citet{2008ApJ...680..580S}; [116]: \citet{2018PASP..130f4002S}; [117]: \citet{2008AJ....136.2306H}; [118]: \citet{2020MNRAS.499..482N}; [119]: \citet{2020MNRAS.497.1925G}; [120]: \citet{2017ApJ...844...46B}; 
    \end{deluxetable*} 
\end{longrotatetable}

\clearpage

\begin{deluxetable*}{cccc}
    \tabletypesize{\scriptsize}
    
    \tablecaption{Reclassifications\label{tab:reclassifications}
}
    \tablehead{
        \colhead{Name} & \colhead{TNS Type} & \colhead{Our Classification} & \colhead{References}
    }

    \startdata
SN2023aew & SNIIb & Transitional & 1 2  \\
SN2020fqv & SNIIb & SNII & 3  \\
SN2022crv & SNIb & SNIIb & 4 5  \\
SN2021gno & SNIb & CaST & 6 7  \\
SN2019ehk & SNIb & CaST & 8 9  \\
SN2008bo & SNIb & SNIIb & 10  \\
SN2010gx & SNIc & Transitional & 11  \\
AT2017gge & Unclassified & TDE & 12  \\
SN2011hw & SNIIn & SNIbn & 13 14  \\
SN2021foa & SNIIn & Transitional & 15 16  \\
SN2009ip & SNIIn & Impostor & 17  \\
SN2018evt & SNIa & SNIa-CSM & 18  \\
SN2008ge & SNIa & SNIax & 19  \\
SN2020aeuh & SNIa & SNIa-CSM & 20  \\
SN2020eyj & SNIa & SNIa-CSM & 21  \\
SN2009dc & SNIa & SNIa-pec & 22  \\
SN2012dn & SNIa & SNIa-pec & 23  \\
SN2022esa & SNIa-CSM & Transitional & 24 25  \\
SN2012Z & SNIa-pec & SNIax & 26 27  \\
SN2005hk & SNIa-pec & SNIax & 28  \\
SN2011ay & SNIa-pec & SNIax & 29  \\
SN2008ae & SNIa-pec & SNIax & 27  \\
SN2007ax & SNIa & SNIa-91bg-like & 30  \\
SN2019yvq & SNIa & SNIa-pec & 31  \\
SN2005ke & SNIa & SNIa-91bg-like & 32  \\
SN2022joj & SNIa & SNIa-pec & 33 34  \\
SN2019hcc & SLSN-I & SLSN-II & 35  \\
SN2020faa & SNII & SNII-pec & 36  \\
SN2018gk & SNII & SNIIb & 37  \\
SN2019tua & SNII & SNIIb & 38  \\
SN2018hna & SNII & SNII-pec & 39  \\
SN2016ezh & SNII & TDE & 40  \\
SN2018bsz & SNII & SLSN-I & 41  \\
AT2017ens & Unclassified & SLSN-I & 42  \\
AT2017gpn & Unclassified & SNIIb & 43  \\
AT2022cmc & Unclassified & TDE & 44  \\
AT2017err & Unclassified & SNIIn & 45  \\
AT2018lqh & Unclassified & FBOT & 46  \\
AT2022tsd & Unclassified & FBOT & 47  \\
AT2021adxl & Unclassified & SNIIn & 48  \\
AT2017gbl & Unclassified & TDE & 49  \\
AT2016bln & Unclassified & SNIa-pec & 50  \\
AT2020xnd & Unclassified & FBOT & 51  \\
AT2016aps & Unclassified & SNIIn & 52  \\
AT2023fhn & Unclassified & FBOT & 53  \\
    \enddata

[1]: \citet{2024AA...689A.182K}; 
[2]: \citet{2024ApJ...966..199S}; 
[3]: \citet{2022MNRAS.512.2777T}; 
[4]: \citet{2023ApJ...957..100G}; 
[5]: \citet{2024ApJ...974..316D}; 
[6]: \citet{2023MNRAS.526..279E}; 
[7]: \citet{2022ApJ...932...58J}; 
[8]: \citet{2020ApJ...898..166J}; 
[9]: \citet{2021ApJ...907L..18D}; 
[10]: \citet{2010ApJ...711L..40C}; 
[11]: \citet{2013ApJ...770..128I}; 
[12]: \citet{2022MNRAS.517...76O}; 
[13]: \citet{2012MNRAS.426.1905S}; 
[14]: \citet{2015MNRAS.449.1921P}; 
[15]: \citet{2022AA...662L..10R}; 
[16]: \citet{2025MNRAS.537.2898G}; 
[17]: \citet{2013ApJ...767....1P}; 
[18]: \citet{2023MNRAS.519.1618Y}; 
[19]: \citet{2010AJ....140.1321F}; 
[20]: \citet{2025AA...704A.135T}; 
[21]: \citet{2023Natur.617..477K}; 
[22]: \citet{2009ApJ...707L.118Y}; 
[23]: \citet{2014MNRAS.443.1663C}; 
[24]: \citet{2025PASJ..tmp..143M}; 
[25]: \citet{2025ApJ...995...54G}; 
[26]: \citet{2015AA...573A...2S}; 
[27]: \citet{2013ApJ...767...57F}; 
[28]: \citet{2014ApJ...786..134M}; 
[29]: \citet{2015MNRAS.453.2103S}; 
[30]: \citet{2008ApJ...683L..29K}; 
[31]: \citet{2021ApJ...919..142B}; 
[32]: \citet{2009ApJ...700.1456B}; 
[33]: \citet{2023ApJ...958..178L}; 
[34]: \citet{2024ApJ...964..196P}; 
[35]: \citet{2021MNRAS.506.4819P}; 
[36]: \citet{2021AA...646A..22Y}; 
[37]: \citet{2021MNRAS.503.3472B}; 
[38]: \citet{2024ApJ...970..103H}; 
[39]: \citet{2019ApJ...882L..15S}; 
[40]: \citet{2017ApJ...843..106B}; 
[41]: \citet{2022AA...666A..30P}; 
[42]: \citet{2018ApJ...867L..31C}; 
[43]: \citet{2021MNRAS.501.5797B}; 
[44]: \citet{2023NatAs...7...88P}; 
[45]: \citet{2022MNRAS.513.4057S}; 
[46]: \citet{2021ApJ...922..247O}; 
[47]: \citet{2025ApJ...993...76O}; 
[48]: \citet{2024AA...690A.259B}; 
[49]: \citet{2020MNRAS.498.2167K}; 
[50]: \citet{2024MNRAS.529.3838A}; 
[51]: \citet{2022ApJ...926..112B}; 
[52]: \citet{2021ApJ...908...99S}; 
[53]: \citet{2024AA...691A.329C};

\end{deluxetable*}

\end{document}